\documentclass[11pt, a4paper]{article}
\pdfoutput=1
\usepackage{jcappub}

\usepackage{booktabs}
\usepackage{multirow}
\usepackage{amssymb}
\usepackage[export]{adjustbox}
\usepackage{floatrow}
\newfloatcommand{capbtabbox}{table}[][3.5in]
\newfloatcommand{capbfigbox}{figure}[][2.3in]
\usepackage{blindtext}
\usepackage{amsmath}
\usepackage{booktabs}
\usepackage{aas_macros}

\newcommand{\ie}{i.e.~}

\def\lsim{\mathrel{\raise.3ex\hbox{$<$\kern-.75em\lower1ex\hbox{$\sim$}}}}
\def\gsim{\mathrel{\raise.3ex\hbox{$>$\kern-.75em\lower1ex\hbox{$\sim$}}}}

\begin{document}

\hspace*{110mm}{\large \tt FERMILAB-PUB-16-245-A}

\vskip 0.2in

\title{The Gamma-Ray Pulsar Population of Globular Clusters: Implications for the GeV Excess}

\author[a,b,c]{Dan Hooper}\note{ORCID: http://orcid.org/0000-0001-8837-4127}
\emailAdd{dhooper@fnal.gov}
\author[d]{and Tim Linden}\note{ORCID: http://orcid.org/0000-0001-9888-0971}
\emailAdd{linden.70@osu.edu}

\affiliation[a]{Fermi National Accelerator Laboratory, Center for Particle
Astrophysics, Batavia, IL 60510}
\affiliation[b]{University of Chicago, Department of Astronomy and Astrophysics, Chicago, IL 60637}
\affiliation[c]{University of Chicago, Kavli Institute for Cosmological Physics, Chicago, IL 60637}
\affiliation[d]{Ohio State University, Center for Cosmology and AstroParticle Physcis (CCAPP), Columbus, OH  43210}

\abstract{It has been suggested that the GeV excess, observed from the region surrounding the Galactic Center, might originate from a population of millisecond pulsars that formed in globular clusters. With this in mind, we employ the publicly available Fermi data to study the gamma-ray emission from 157 globular clusters, identifying a statistically significant signal from 25 of these sources (ten of which are not found in existing gamma-ray catalogs). We combine these observations with the predicted pulsar formation rate based on the stellar encounter rate of each globular cluster to constrain the gamma-ray luminosity function of millisecond pulsars in the Milky Way's globular cluster system. We find that this pulsar population exhibits a luminosity function that is quite similar to those millisecond pulsars observed in the field of the Milky Way ({\ie}the thick disk). After pulsars are expelled from a globular cluster, however, they continue to lose rotational kinetic energy and become less luminous, causing their luminosity function to depart from the steady-state distribution. Using this luminosity function and a model for the globular cluster disruption rate, we show that millisecond pulsars born in globular clusters can account for only a few percent or less of the observed GeV excess. Among other challenges, scenarios in which the entire GeV excess is generated from such pulsars are in conflict with the observed mass of the Milky Way's Central Stellar Cluster.}

\maketitle

\section{Introduction}

The Fermi Gamma-Ray Space Telescope has detected gamma-ray emission from a number of globular clusters~\cite{collaboration:2010bb,TheFermi-LAT:2015hja,Cholis:2014noa}, generally thought to originate from the millisecond pulsars (MSPs) that reside within these systems~\cite{Harding:2004hj,Bednarek:2007nn,Venter:2009mq,Cheng:2010za,collaboration:2010bb,msps_47Tuc,msps_10_47Tuc, msps_more_47Tuc, Camilo:1999fc,Bogdanov:2006ap}. The detection of pulsed gamma-ray emission from two globular clusters~\cite{2011Sci...334.1107F,TheFermi-LAT:2013ssa,Johnson:2013uza} has further strengthened the case for this connection. By studying globular clusters at gamma-ray and other wavelengths, one can learn about the evolution of these systems, as well as the mechanisms that lead to the formation of MSPs in globular cluster environments.

Although Fermi has detected gamma-ray emission from 205 pulsars, the vast majority of these sources are not located within a globular cluster~\cite{2009Sci...325..848A,TheFermi-LAT:2013ssa}.\footnote{For an updated list of Fermi-detected pulsars, see \url{https://confluence.slac.stanford.edu/display/GLAMCOG/Public+List+of+LAT-Detected+Gamma-Ray+Pulsars}} Of these pulsars, 93 exhibit millisecond-scale periods~\cite{TheFermi-LAT:2013ssa,Guillemot:2011th,2009Sci...325..848A}. Such MSPs are thought to represent a distinct population of sources, which form through the interaction of an old neutron star with a stellar companion~\cite{1982Natur.300..728A,1994ARA&A..32..591P,Lorimer:2001vd,Lorimer:2008se,2010ApJ...715..335K}. In contrast to other pulsars, MSPs have much weaker magnetic fields and thus lose their rotational kinetic energy more slowly, remaining luminous for as long as billions of years.  Unlike pulsar populations found elsewhere in the Galaxy, those in globular clusters consist almost entirely of MSPs, enhanced by the high stellar densities and stellar encounter rates of such systems.

A bright and statistically significant gamma-ray signal has been detected from the region surrounding the Galactic Center~\cite{Goodenough:2009gk,Hooper:2010mq,Hooper:2011ti,Abazajian:2012pn,Gordon:2013vta,Hooper:2013rwa,Daylan:2014rsa,Calore:2014xka,TheFermi-LAT:2015kwa}, with angular and spectral features that are consistent with those predicted from annihilating dark matter particles. And while this possibility has generated a great deal of interest (see, for example, Refs.~\cite{Abdullah:2014lla,Ipek:2014gua,Izaguirre:2014vva,Agrawal:2014una,Berlin:2014tja,Alves:2014yha,Boehm:2014hva,Martin:2014sxa,Huang:2014cla,Cerdeno:2014cda,Okada:2013bna,Freese:2015ysa,Fonseca:2015rwa,Bertone:2015tza,Cline:2015qha,Berlin:2015wwa,Caron:2015wda,Cerdeno:2015ega,Liu:2014cma,Hooper:2014fda,Arcadi:2014lta,Cahill-Rowley:2014ora,Ko:2014loa,McDermott:2014rqa}), astrophysical explanations have also been proposed. The leading astrophysical interpretation of the GeV excess is that it is generated by a large population of unresolved gamma-ray emitting MSPs, densely concentrated in a spherically symmetric distribution around the Galactic Center~\cite{Hooper:2010mq,Hooper:2011ti,Abazajian:2012pn,Gordon:2013vta,Yuan:2014rca,Petrovic:2014xra,Brandt:2015ula}. This possibility has been motivated in large part by the measured spectral shape of this signal, which is similar to that observed from MSPs. Recent analyses have also reported tentative evidence of an unresolved point source population in this region of the sky~\cite{Lee:2015fea,Bartels:2015aea}. It is not yet clear, however, whether these studies have identified signatures of a sub-threshold point source population, or merely variations in the gamma-ray flux associated with the small scale structure of the diffuse background (see, for example, Ref.~\cite{Horiuchi:2016zwu}). 

Several arguments have been made against the possibility that MSPs could account for the observed excess emission. First, if this signal originates from MSPs, then one should expect Fermi to have detected and resolved many more bright gamma-ray point sources from the direction of the Inner Galaxy than have been reported (assuming that the luminosity function of MSPs in the Inner Galaxy is similar to that of other MSP populations)~\cite{Hooper:2015jlu,Cholis:2014lta,SiegalGaskins:2010mp}. Second, such an MSP population in the Inner Galaxy should be accompanied by approximately 20 times as many bright low-mass X-ray binaries (LMXBs) than are observed~\cite{Cholis:2014lta}. And third, the spherical morphology of the observed excess does not resemble the distribution of any known stellar population in the Galactic Bulge. Instead, such populations are less spatially concentrated and more extended along the direction of the Galactic Disk than is exhibited by the excess.

It is plausible that all three of these problems could be circumvented, or at least mitigated, in a scenario in which the Inner Galaxy's MSP population originates from a large number of tidally disrupted globular clusters~\cite{Brandt:2015ula,Bednarek:2013oha}. Over Gyr timescales, dynamical friction can cause globular clusters to spiral into the Inner Galaxy, where they are tidally disrupted~\cite{Gnedin:2013cda,2012ApJ...750..111A,2013ApJ...763...62A}. When a globular cluster is destroyed in this way, their stars (including MSPs) are deposited into the Galactic Bulge and/or Central Stellar Cluster. As new MSPs are no longer efficiently generated after this point in this time, the deposited pulsar population evolves, losing rotational kinetic energy and luminosity via magnetic dipole breaking. This evolution could potentially explain both the lack of bright pulsars detections, as well as the lack of bright LMXBs. Although no simulations using a realistic 3D potential have been carried out to date, it may be possible that the distribution of such MSPs could be approximately spherical, at least to first order. 

In this paper, we revisit the population of gamma-ray emitting millisecond pulsars (MSPs) that are present within globular clusters. We study the publicly available Fermi data from the direction of 157 globular clusters in an effort to detect and characterize the gamma-ray emission from these sources. We identify 25 of these globular clusters as statistically significant gamma-ray sources (TS $>$ 25), ten of which are not contained in previous gamma-ray source catalogs. We also report low-significance detections of many other globular clusters, suggesting that a large fraction of this population may be gamma-ray emitting. We use this data, in conjunction with previously calculated stellar encounter rates, to constrain the gamma-ray luminosity function of MSPs within globular clusters. We find that this luminosity function is similar to that reported previously for MSPs in the field ({\it i.e.}~the thick disk) of the Milky Way. We then consider the effects of spin-down evolution on the MSP luminosity function, for a population deposited into the Inner Galaxy following the tidal disruption of their parent clusters. We estimate that MSPs within this population will be approximately an order of magnitude less luminous on average than those found within intact globular clusters, and that such pulsars are likely to generate a total gamma-ray flux that is much smaller (a few percent or less) than that of the observed GeV excess. Although one could plausibly construct models in which a larger number of globular clusters have been tidally disrupted in the Inner Galaxy, such models cannot accommodate the observed intensity of the GeV excess without exceeding the mass of the Milky Way's Central Stellar Cluster.


\section{Gamma-Rays From Globular Clusters}
\label{gamma}

The gamma-ray emission observed from globular clusters~\cite{collaboration:2010bb,TheFermi-LAT:2015hja,Cholis:2014noa} is generally assumed to originate largely from MSPs that are contained within those clusters~\cite{collaboration:2010bb,msps_47Tuc, msps_10_47Tuc, msps_more_47Tuc, Camilo:1999fc,Bogdanov:2006ap}. Globular clusters generally appear as point sources to Fermi, and the most recent Fermi Source Catalog (the 3FGL)~\cite{TheFermi-LAT:2015hja} includes 15 sources that are associated with a globular cluster (see also Ref.~\cite{Cholis:2014noa}). 

In this study, we have analyzed the publicly available Fermi data in the directions of 157 globular clusters (approximately the entirety of the Milky Way's globular cluster population). For each of these 157 sources, we have determined the intensity and spectrum of their gamma-ray emission, using 85 months of Fermi-LAT data.\footnote{MET Range: 239557417 --- 464084557} We have restricted our analysis to events that meet the Pass 8 Source event selection criteria. We have applied standard cuts to the data, removing events that were recorded at a zenith angle larger than 90$^\circ$, while the instrument is not in Survey mode, while the instrumental rocking angle exceeds 52$^\circ$, and while Fermi was passing through the South Atlantic Anomaly. Given the low luminosities of many globular clusters, we have placed no constraints on the point spread function class and include both front and back converting events.

For each globular cluster, we divide the resulting data set into 15 logarithmic energy bins between 0.1 and 100 GeV, as well as 280$\times$280 angular bins spanning a 14$^\circ\times$14$^\circ$ region-of-interest centered on the position of the source. We then fit the normalization and spectrum of each source according to the following procedure. First, we fit all background components over the full sky and over the full energy range using a spectral model for each source. In this stage we include the full 3FGL point source catalog~\cite{TheFermi-LAT:2015hja} and the {\tt gll\_iem\_v06.fits} Fermi diffuse emission model, as recommended by the Fermi Collaboration for Pass 8 data. We also employ the matching isotropic background model {\tt iso\_P8R2\_SOURCE\_V6\_v06.txt}. We use the standard Fermi-LAT algorithm to determine whether a given source component should be allowed to float freely, or be held fixed in the fit, and use the python implementation of the {\tt gtlike} tool, including the {\tt MINUIT} algorithm, to determine the best-fit normalization and spectrum of each emission component.\footnote{For the small fraction of globular clusters that are located close to many 3FGL sources, we fix the spectra of the most distant sources until the number of degrees-of-freedom associated with the point source backgrounds is reduced to less than 100 (the spectra of all sources within 4$^{\circ}$ of a given globular cluster are always allowed to float).} Then, we add a point source at the position of the globular cluster, and perform a full scan of the likelihood fit as a function of the flux, independently in each energy bin. We fit the resulting likelihood distribution assuming an arbitrarily normalized spectra of the form of a power-law with an exponential cut-off,  ${dN_{\gamma}/dE_{\gamma} \propto E_{\gamma}^{-\alpha}} \, e^{-E_{\gamma}/E_{\rm cut}}$, allowing $\alpha$ and $E_{\rm cut}$ to float between 0 and 3.5 and 0.1 and 100 GeV, respectively. By determining the maximum improvement to the likelihood, we calculate the value of the test statistic (TS) for each globular cluster, as well as the likelihood profiles for the gamma-ray flux and spectral index.

\begin{table}
\renewcommand{\arraystretch}{1.2}
\begin{tabular}{|c|c|c|c|c|c|}
\hline 
Globular Cluster & Alternate Name & Flux (erg/cm$^2$/s) & $\alpha$ & $E_{\rm cut}$(GeV) & TS \tabularnewline
\hline 
\hline 
NGC 104 & 47 Tuc  & $    2.436^{+0.062}_{-0.062} \times 10^{-11}$ & 1.18 &  2.51 & 4055.9 \tabularnewline
 \hline
NGC 2808 &          & $    3.546^{+ 0.602}_{- 0.486} \times 10^{-12}$ & 1.36 &  3.16 &   97.4 \tabularnewline
 \hline
 NGC 5139 & Omega Centauri     & $    5.900^{+ 0.468}_{- 0.453} \times 10^{-12}$ &-0.12 &  1.26 &  301.3 \tabularnewline
 \hline
   NGC 5904          & M5 & $    2.131^{+ 0.539}_{- 0.600} \times 10^{-12}$ & 1.86 &  3.98 &   39.6 \tabularnewline
 \hline
 NGC 6093 & M80     & $    3.986^{+ 0.596}_{- 0.705} \times 10^{-12}$ & 1.38 &  5.01 &   96.9 \tabularnewline
 \hline
   NGC 6139          &  & $    5.330^{+ 1.310}_{- 0.936} \times 10^{-12}$ & 2.28 & 19.95 &   40.6 \tabularnewline
 \hline
   NGC 6218          &  M12& $    2.969^{+ 0.655}_{- 0.844} \times 10^{-12}$ & 2.24 & $\ge 100$ &   31.0 \tabularnewline
 \hline
 NGC 6266 &  M62 & $    1.710^{+ 0.074}_{- 0.070} \times 10^{-11}$ & 1.36 &  3.16 &  855.7 \tabularnewline
 \hline
NGC 6316 &   & $    1.091^{+ 0.124}_{- 0.120} \times 10^{-11}$ & 2.00 &  7.94 &  163.5 \tabularnewline
 \hline
  NGC 6342          &  & $    4.339^{+ 1.046}_{- 1.015} \times 10^{-12}$ & 2.16 & 15.85 &   37.8 \tabularnewline
 \hline
  NGC 6388 &      & $    1.732^{+ 0.124}_{- 0.099} \times 10^{-11}$ & 1.52 &  3.16 &  779.6 \tabularnewline
 \hline
  NGC 6397          &  & $    6.390^{+ 0.734}_{- 0.727} \times 10^{-12}$ & 2.90 & 50.12 &   81.5 \tabularnewline
 \hline
   Palomar 6             &  & $    5.489^{+ 1.455}_{- 1.324} \times 10^{-12}$ & 0.94 &  1.26 &   29.9 \tabularnewline
 \hline
  Terzan 5 & Terzan 11  & $    5.973^{+ 0.203}_{- 0.147} \times 10^{-11}$ & 1.16 &  2.51 & 2742.3 \tabularnewline
 \hline
 NGC 6440 &        & $    2.392^{+ 0.178}_{- 0.105} \times 10^{-11}$ & 2.32 & 10.00 &  390.6 \tabularnewline
 \hline
NGC 6441 &   & $    1.252^{+ 0.088}_{- 0.144} \times 10^{-11}$ & 2.04 & 10.00 &  217.8 \tabularnewline
 \hline
NGC 6541 &      & $    3.251^{+ 0.748}_{- 0.667} \times 10^{-12}$ & 1.16 &  2.51 &   77.7 \tabularnewline
 \hline 
2MASS-GC01 &     & $    2.476^{+ 0.217}_{- 0.196} \times 10^{-11}$ & 1.06 &  1.26 &  179.8 \tabularnewline
 \hline
  2MASS-GC02        &  & $    8.846^{+ 2.051}_{- 2.065} \times 10^{-12}$ & 1.08 &  1.26 &   28.2 \tabularnewline
 \hline
 GLIMPSE 02        &  & $    1.630^{+ 0.228}_{- 0.242} \times 10^{-11}$ & 1.94 &  7.94 &   67.3 \tabularnewline
 \hline
 NGC 6652 &      & $    4.495^{+ 0.805}_{- 0.495} \times 10^{-12}$ & 1.38 &  3.16 &  128.5 \tabularnewline
 \hline
   GLIMPSE 01        &  & $    9.020^{+ 1.205}_{- 1.345} \times 10^{-12}$ &-0.74 &  1.58 &   68.7 \tabularnewline
 \hline
 NGC 6717 & Palomar 9    & $    1.816^{+ 0.543}_{- 0.386} \times 10^{-12}$ & 0.38 &  2.51 &   42.3 \tabularnewline
 \hline
 NGC 6752 &    & $    2.866^{+ 0.503}_{- 0.327} \times 10^{-12}$ & 0.12 &  0.79 &  144.8 \tabularnewline
 \hline
  NGC 7078          & M15 & $    3.160^{+ 0.587}_{- 0.604} \times 10^{-12}$ & 2.42 &  6.31 &   41.8 \tabularnewline
 \hline
  \end{tabular}
\caption{The gamma-ray flux (0.1-100 GeV) and best-fit spectral parameters for the 25 globular clusters detected with TS $>$ 25 in our analysis of the Fermi data.}
\label{table1}
\end{table}

Our analysis identified statistically significant (TS $>$ 25) gamma-ray emission from 25 globular clusters, 15 of which are found in the 3FGL catalog. In Table~\ref{table1}, we list the gamma-ray flux (integrated between 0.1 and 100 GeV) and best-fit spectral parameters for each of these 25 sources (see Table~\ref{table2} for the 3FGL names of these globular clusters). The errors quoted for the fluxes in this table denote the 1$\sigma$ range, as determined using the full 3D likelihood profile. As expected, most of these sources exhibit spectra that peak at energies near $\sim$1-2 GeV (in $E_{\gamma}^2 dN_{\gamma}/dE_{\gamma}$ units), further supporting a pulsar interpretation.

As a check, we have compared our results to the information provided in the 3FGL catalog, finding that our fluxes are systematically lower than those reported by the Fermi Collaboration (in most cases by $\sim$20-50\%). Upon further investigation, we determined that this variation is largely the result of the spectral parameterizations adopted in these fits. More specifically, whereas we have adopted an exponentially cutoff power-law form, the Fermi Collaboration restricts the spectra of globular clusters to be fit by either a power-law (without any cutoff), or a log-parabola. When we perform our fits using these spectral parameterizations, our fluxes are in much better agreement with those listed in the 3FGL. Given the gamma-ray spectra observed from MSPs, our choice of spectral parameterization is well motivated, and in most cases provides a better fit to the data.

\begin{table}
\renewcommand{\arraystretch}{1.2}
\begin{tabular}{|c|c|}
\hline 
Globular Cluster & 3FGL Name  \tabularnewline
\hline 
\hline 
NGC 104 & 3FGL J0023.9-7203   \tabularnewline
 \hline
NGC 2808 & 3FGL J0912.2-6452       \tabularnewline
 \hline
 NGC 5139 &  3FGL J1326.7-4727   \tabularnewline
 \hline
  NGC 6093 & 3FGL J1616.8-2300    \tabularnewline
 \hline
 NGC 6266 &   3FGL J1701.2-3006 \tabularnewline
 \hline
NGC 6316 & 3FGL J1716.6-2812   \tabularnewline
 \hline
    NGC 6388 & 3FGL  J1736.2-4444   \tabularnewline
\hline
  Terzan 5 &  3FGL J1748.0-2447 \tabularnewline
 \hline
 NGC 6440 & 3FGL J1748.9-2021   \tabularnewline
 \hline
NGC 6441 & 3FGL J1750.2-3704   \tabularnewline
 \hline
NGC 6541 & 3FGL J1807.5-4343   \tabularnewline
 \hline 
2MASS-GC01 & 3FGL J1808.5-1952  \tabularnewline
 \hline
  NGC 6652 & 3FGL J1835.7-3258   \tabularnewline
 \hline
   NGC 6717 &  3FGL J1855.1-2243  \tabularnewline
 \hline
   NGC 6752 & 3FGL J1910.7-6000   \tabularnewline
 \hline
  \end{tabular}
\caption{The names of the 15 globular clusters included in the 3FGL catalog.}
\label{table2}
\end{table}

\begin{table}
\renewcommand{\arraystretch}{1.2}
\begin{tabular}{|c|c|c|c|}
\hline 
Globular Cluster & Alternate Name & Flux (erg/cm$^2$/s) &  TS \tabularnewline
\hline 
\hline 
  Palomar 5             &  & $    <1.56 \times 10^{-12}$ &  28.7 \tabularnewline
 \hline
   Lynga 7           & BH 184 & $ <   4.96 \times 10^{-12}$ &   31.9 \tabularnewline
   \hline
  NGC 6205          & M13 & $   < 1.54 \times 10^{-12}$ &   26.8 \tabularnewline
 \hline
   Terzan 1          & HP 2 & $  <  4.95 \times 10^{-12}$ &  76.4 \tabularnewline
 \hline
    Ton 2             &  Pismis 26 & $ <   3.99 \times 10^{-12}$ &   25.0 \tabularnewline
 \hline
    NGC 6401          &  & $   < 5.50 \times 10^{-12}$ &   74.9 \tabularnewline
 \hline
  NGC 6535          &  & $  <  4.44 \times 10^{-12}$ &   55.6 \tabularnewline
 \hline
  IC 1276             &  & $   <  8.25 \times 10^{-12}$ &    72.2 \tabularnewline
 \hline
  NGC 6569          &  & $   < 1.10 \times 10^{-12}$ &   33.7 \tabularnewline
 \hline
    NGC 6712          &  & $  <  8.44  \times 10^{-12}$ &    96.9 \tabularnewline
 \hline
    NGC 6749          &  & $   < 4.70 \times 10^{-12}$ &   27.4 \tabularnewline
 \hline
  \end{tabular}
\caption{The 2$\sigma$ upper limits on the gamma-ray flux (0.1-100 GeV) for those globular clusters detected by Fermi with TS $>$ 25, but only for spectra that peak at energies below 0.4 GeV.}
\label{table3}
\end{table}

In addition to these 25 sources, there are 11 others that were found to yield detections at the level of TS $>$ 25, but with little or no statistical significance above 0.4 GeV. Given Fermi's poor point spread function at such low energies, we do not treat these as unambiguous detections, and instead report only an upper limit on their gamma-ray flux. We also note that the majority of these 11 sources are located within 5$^{\circ}$ of the Galactic Plane, where Fermi data analysis is most challenging. These 11 sources and the upper limits on their fluxes are listed in Table~\ref{table3}.

In Tables~\ref{table10}-\ref{table12}, we provide upper limits on the gamma-ray fluxes from those globular clusters that were not detected with high significance in our analysis. We note that although these sources each yielded TS $<$ 25, a sizable number have marginal detections. 
%
%
%
We consider it likely than many of these globular clusters are in fact gamma-ray emitters, with the potential to be detected with more data from Fermi, or with a post-Fermi satellite mission. On the other hand, we expect some false detections, in particular among clusters located near the Galactic Plane.\footnote{Using a ``mirrored'' sky location test following Ref.~\cite{Linden:2015qha}, we anticipate a false detection rate from our globular cluster
population of 5.8 systems with TS $>$ 25 and 2.5 systems with TS $>$ 30, using the same spectral cut employed for our globular cluster population.}

\begin{table}
\begin{tabular}{|c|c|c|c|}
\hline 
Globular Cluster & Alternate Name & Flux (erg/cm$^2$/s) & TS  \tabularnewline
\hline 
\hline 
NGC 288             &  & $<   6.45 \times 10^{-13}$ & 2.94 \tabularnewline
 \hline
 NGC 362             &  & $<   8.91 \times 10^{-13}$ &14.64 \tabularnewline
 \hline 
   Whiting 1         &  & $<   1.48 \times 10^{-12}$ & 8.11 \tabularnewline
 \hline
 NGC 1261            &  & $<   1.50 \times 10^{-12}$ & 5.70 \tabularnewline
 \hline
   Palomar 1             &  & $<   1.64 \times 10^{-12}$ & 6.51 \tabularnewline
 \hline
  AM-1 & E1 & $< 7.97 \times 10^{-13}$ &  0.32 \tabularnewline
\hline
  Eridanus            &  & $<   9.19 \times 10^{-13}$ & 1.63 \tabularnewline
 \hline
  Palomar 2             &  & $<   1.42 \times 10^{-12}$ & 0.00 \tabularnewline
 \hline
  NGC 1851            &  & $<   1.69 \times 10^{-12}$ &18.27 \tabularnewline
 \hline
 NGC 1904            & M79 & $<   2.32 \times 10^{-12}$ &15.03 \tabularnewline
 \hline
 NGC 2298            &  & $<   2.16 \times 10^{-12}$ & 4.47 \tabularnewline
 \hline
 NGC 2419            &  & $<   1.31 \times 10^{-12}$ & 2.16 \tabularnewline
 \hline
 Ko 2                &  & $<   6.31 \times 10^{-13}$ & 0.08 \tabularnewline
 \hline
  Pyxis             &  & $<   1.67 \times 10^{-12}$ & 0.70 \tabularnewline
 \hline
 E3                 &  & $<   2.04 \times 10^{-12}$ & 7.17 \tabularnewline
 \hline
   Palomar 3             &  & $<   7.06 \times 10^{-13}$ & 3.76 \tabularnewline
 \hline
 NGC 3201            &  & $<   1.99 \times 10^{-12}$ & 7.80 \tabularnewline
 \hline
   Palomar 4             &  & $<   2.00 \times 10^{-12}$ & 8.20 \tabularnewline
 \hline
  Ko 1                &  & $<   7.31 \times 10^{-13}$ & 0.77 \tabularnewline
 \hline
  NGC 4147            &  & $<   8.16 \times 10^{-13}$ & 0.75 \tabularnewline
 \hline
 NGC 4372            &  & $<   3.00 \times 10^{-12}$ &17.29 \tabularnewline
 \hline
   Rup 106           &  & $<   2.32 \times 10^{-12}$ & 5.61 \tabularnewline
 \hline
 NGC 4590            &  M68 & $<   2.65 \times 10^{-12}$ &11.84 \tabularnewline
 \hline
 NGC 4833            &  & $<   2.42 \times 10^{-12}$ & 2.43 \tabularnewline
 \hline
 NGC 5024            & M53 & $<   2.91 \times 10^{-12}$ &21.29 \tabularnewline
 \hline
 NGC 5053            &  & $<   2.73 \times 10^{-12}$ &17.23 \tabularnewline
 \hline
   NGC 5272          & M3   & $<   1.92 \times 10^{-12}$ & 6.92 \tabularnewline
 \hline
 NGC 5286          &  & $<   4.07 \times 10^{-12}$ &13.79 \tabularnewline
 \hline
 AM-4 & & $<1.60 \times 10^{-12}$ &   3.06 \tabularnewline
\hline
 NGC 5466          &  & $<   1.12 \times 10^{-12}$ & 1.04 \tabularnewline
 \hline
  NGC 5634          &  & $<   1.08 \times 10^{-12}$ & 0.43 \tabularnewline
 \hline
  NGC 5694          &  & $<   1.04 \times 10^{-12}$ & 3.75 \tabularnewline
 \hline
  IC 4499             &  & $<   3.20 \times 10^{-12}$ & 23.49 \tabularnewline
 \hline
   NGC 5824          &  & $<   3.97 \times 10^{-13}$ & 0.00 \tabularnewline
 \hline
   NGC 5897          &  & $<   2.50 \times 10^{-12}$ & 5.32 \tabularnewline
 \hline
   NGC 5927          &  & $<   3.55 \times 10^{-12}$ & 3.12 \tabularnewline
 \hline
   NGC 5946          &  & $<   3.14 \times 10^{-12}$ & 1.14 \tabularnewline
 \hline
  BH 176            &  & $<   2.69 \times 10^{-12}$ & 3.99 \tabularnewline
 \hline
   NGC 5986          &  & $<   3.95 \times 10^{-12}$ &17.05 \tabularnewline
 \hline
   Palomar 14            & AvdB & $<   5.61 \times 10^{-13}$ & 0.14 \tabularnewline
 \hline
 \end{tabular}
\caption{The 2$\sigma$ upper limits on the gamma-ray flux (0.1-100 GeV) for those globular clusters that were not significantly detected by Fermi (TS $<$ 25).}
\label{table10}
\end{table}

\begin{table}
\begin{tabular}{|c|c|c|c|}
\hline 
Globular Cluster & Alternate Name & Flux (erg/cm$^2$/s) & TS  \tabularnewline
\hline 
\hline
   NGC 6121          & M4  & $<   4.45 \times 10^{-12}$ &17.57 \tabularnewline
\hline
   NGC 6101          &  & $<   1.72 \times 10^{-12}$ &12.65 \tabularnewline
 \hline
  NGC 6144          &  & $<   3.03 \times 10^{-12}$ & 2.75 \tabularnewline
 \hline
    Terzan 3          &  & $<   3.64 \times 10^{-12}$ & 5.59 \tabularnewline
 \hline
   NGC 6171          & M107 & $<   2.43 \times 10^{-12}$ & 2.19 \tabularnewline
   \hline
   GC 1636-283        &  ESO 0452-SC11 & $<   3.84 \times 10^{-12}$ &17.09 \tabularnewline
 \hline
  NGC 6229          &  & $<   7.64 \times 10^{-13}$ & 0.00 \tabularnewline
 \hline
   FSR 1735         &  & $<   4.63 \times 10^{-12}$ &12.94 \tabularnewline
 \hline
   NGC 6235          &  & $<   2.21 \times 10^{-12}$ & 1.49 \tabularnewline
 \hline
   NGC 6254          & M10 & $<   2.14 \times 10^{-12}$ &18.39 \tabularnewline
 \hline
   NGC 6256          &  & $<   5.80 \times 10^{-12}$ &10.55 \tabularnewline
 \hline
   Palomar 15            &  & $<   2.34 \times 10^{-12}$ & 3.25 \tabularnewline
 \hline
   NGC 6273          & M19 & $<   1.16 \times 10^{-12}$ & 3.69 \tabularnewline
 \hline
  NGC 6284          &  & $<   2.44 \times 10^{-12}$ & 2.09 \tabularnewline
 \hline
   NGC 6287          &  & $<   4.41 \times 10^{-12}$ & 8.88 \tabularnewline
 \hline
   NGC 6293          &  & $<   3.24 \times 10^{-12}$ & 4.74 \tabularnewline
 \hline
   NGC 6304          &  & $<   5.98 \times 10^{-12}$ &16.05 \tabularnewline
 \hline
   NGC 6341          & M92 & $<   1.51 \times 10^{-12}$ &11.86 \tabularnewline
 \hline
  NGC 6325          &  & $<   2.35 \times 10^{-12}$ & 0.53 \tabularnewline
 \hline
   NGC 6333          &  M9& $<   1.76 \times 10^{-12}$ & 3.94 \tabularnewline
 \hline
  NGC 6356          &  & $<   1.29 \times 10^{-12}$ & 0.08 \tabularnewline
 \hline
 NGC 6355          &  & $<   2.67 \times 10^{-12}$ & 0.53 \tabularnewline
 \hline 
   NGC 6352          &  & $<   6.17 \times 10^{-12}$ &21.74 \tabularnewline
 \hline
  IC 1257             &  & $<   2.88 \times 10^{-12}$ & 5.74 \tabularnewline
 \hline
   Terzan 2          & HP 3 & $<   2.96 \times 10^{-12}$ &16.38 \tabularnewline
 \hline
   NGC 6366          &  & $<   3.23 \times 10^{-12}$ & 5.36 \tabularnewline
 \hline
   Terzan 4          & HP 4 & $<   4.10 \times 10^{-12}$ & 0.88 \tabularnewline
 \hline
HP 1              & BH 229 & $<   4.79 \times 10^{-12}$ & 6.98 \tabularnewline
 \hline
   NGC 6362          &  & $<   1.71 \times 10^{-12}$ & 2.31 \tabularnewline
 \hline
    Liller 1          &  & $<   4.95 \times 10^{-12}$ & 5.28 \tabularnewline
 \hline
  NGC 6380          & Ton 1  & $<   6.51 \times 10^{-12}$ &11.57 \tabularnewline
 \hline
   NGC 6402          & M14 & $<   2.29 \times 10^{-12}$ &24.64 \tabularnewline
 \hline
   NGC 6426          &  & $<   2.26 \times 10^{-12}$ & 2.57 \tabularnewline
 \hline
   Djorg 1           &  & $<   2.86 \times 10^{-12}$ & 1.05 \tabularnewline
 \hline
   Terzan 6          &  HP 5 & $<   5.46 \times 10^{-12}$ & 5.26 \tabularnewline
 \hline
   NGC 6453          &  & $<   1.06 \times 10^{-12}$ & 0.00 \tabularnewline
 \hline
  UKS 1            &  & $<   4.68 \times 10^{-12}$ & 1.49 \tabularnewline
 \hline
   NGC 6496          &  & $<   3.06 \times 10^{-12}$ & 8.18 \tabularnewline
 \hline
   Terzan 9          &  & $<   6.86 \times 10^{-12}$ &10.82 \tabularnewline
 \hline
   Djorg 2           & ESO 456-SC38 & $<   5.56 \times 10^{-12}$ & 9.50 \tabularnewline
 \hline
 \end{tabular}
\caption{Continued from Table~\ref{table10}. The 2$\sigma$ upper limits on the gamma-ray flux (0.1-100 GeV) for those globular clusters that were not significantly detected by Fermi (TS $<$ 25).}
\label{table11}
\end{table}

\begin{table}
\begin{tabular}{|c|c|c|c|}
\hline 
Globular Cluster & Alternate Name & Flux (erg/cm$^2$/s) & TS  \tabularnewline
\hline
   NGC 6517          &  & $<   2.11 \times 10^{-12}$ &12.72 \tabularnewline
 \hline
   Terzan 10         &  & $<   5.06 \times 10^{-12}$ & 3.90 \tabularnewline
 \hline
   NGC 6522          &  & $<   2.19 \times 10^{-12}$ & 1.49 \tabularnewline
\hline
   NGC 6528         &  & $<   3.34 \times 10^{-12}$ & 3.73 \tabularnewline
 \hline
  NGC 6539          &  & $<   7.61 \times 10^{-12}$ &20.47 \tabularnewline
 \hline
   NGC 6540          & Djorg 3 & $<   5.45 \times 10^{-12}$ & 8.99 \tabularnewline
 \hline
   NGC 6544          &  & $<   2.84 \times 10^{-12}$ & 5.41 \tabularnewline
 \hline
  ESO-SC06            &  ESO280-SC06 & $<   4.00 \times 10^{-12}$ &19.57 \tabularnewline
 \hline
   NGC 6553          &  & $<   1.63 \times 10^{-12}$ & 0.00 \tabularnewline
 \hline
   NGC 6558          &  & $<   2.92 \times 10^{-12}$ & 2.19 \tabularnewline
 \hline
   Terzan 12         &  & $<   2.88 \times 10^{-12}$ & 0.02 \tabularnewline
 \hline
  BH 261            & AL 3 & $<   1.64 \times 10^{-12}$ & 0.00 \tabularnewline
 \hline
   NGC 6584          &  & $<   2.48 \times 10^{-12}$ & 9.01 \tabularnewline
\hline 
    NGC 6624          &  & $<   2.77 \times 10^{-12}$ &16.02 \tabularnewline
 \hline
   NGC 6626          & M28 & $<   4.02 \times 10^{-12}$ & 5.89 \tabularnewline
 \hline
   NGC 6638          &  & $<   5.44 \times 10^{-12}$ &19.30 \tabularnewline
 \hline
   NGC 6637          & M69 & $<   1.86 \times 10^{-12}$ &11.99 \tabularnewline
 \hline
   NGC 6642          &  & $<   3.48 \times 10^{-12}$ & 3.70 \tabularnewline
 \hline
  NGC 6656          &  M22 & $<   2.30 \times 10^{-12}$ &11.00 \tabularnewline
 \hline
     Palomar 8             &  & $<   4.29 \times 10^{-12}$ & 8.76 \tabularnewline
 \hline
   NGC 6681          & M70  & $<   1.83 \times 10^{-12}$ &10.10 \tabularnewline
 \hline
  NGC 6715          & M54 & $<   1.76 \times 10^{-12}$ & 5.35 \tabularnewline
 \hline
 NGC 6723          &  & $<   1.94 \times 10^{-12}$ & 3.92 \tabularnewline
 \hline
   NGC 6760          &  & $<   7.92 \times 10^{-12}$ &19.79 \tabularnewline
 \hline
   NGC 6779          & M56 & $<   2.72 \times 10^{-12}$ & 8.32 \tabularnewline
 \hline
   Terzan 7          &  & $<   2.35 \times 10^{-12}$ & 8.21 \tabularnewline
 \hline
   Palomar 10            &  & $<   8.62 \times 10^{-12}$ &23.62 \tabularnewline
 \hline
 Arp 2               &  & $<   1.83 \times 10^{-12}$ & 1.41 \tabularnewline
 \hline
  NGC 6809          & M55 & $<   2.42 \times 10^{-12}$ & 5.26 \tabularnewline
 \hline
   Terzan 8          &  & $<   2.43 \times 10^{-12}$ & 4.95 \tabularnewline
 \hline
   Palomar 11            &  & $<   1.66 \times 10^{-12}$ & 4.46 \tabularnewline
 \hline
 NGC 6838          & M71 & $<   5.96 \times 10^{-12}$ &24.47 \tabularnewline
 \hline
   NGC 6864          &  M75 & $<   1.11 \times 10^{-12}$ & 1.37 \tabularnewline
 \hline
  NGC 6934          &  & $<   1.51 \times 10^{-12}$ & 1.46 \tabularnewline
 \hline
  NGC 6981          &  M72 & $<   1.38 \times 10^{-12}$ & 1.27 \tabularnewline
 \hline
  NGC 7006          &  & $<   1.94 \times 10^{-12}$ & 3.78 \tabularnewline
 \hline  
  NGC 7089           &  M2 & $<   9.42 \times 10^{-13}$ & 0.74 \tabularnewline
 \hline
NGC 7099          &  M30 & $<   1.32 \times 10^{-12}$ & 8.39 \tabularnewline
 \hline
   Palomar 12            &  & $<   9.53 \times 10^{-13}$ & 0.23 \tabularnewline
 \hline
  Palomar 13            &  & $<   3.01 \times 10^{-12}$ & 7.45 \tabularnewline
 \hline
 NGC 7492        &  & $<   1.17 \times 10^{-12}$ & 4.07 \tabularnewline
 \hline
  \end{tabular}
\caption{Continued from Tables~\ref{table10} and~\ref{table11}. The 2$\sigma$ upper limits on the gamma-ray flux (0.1-100 GeV) for those globular clusters that were not significantly detected by Fermi (TS $<$ 25).}
\label{table12}
\end{table}


\section{The Luminosity Function of Pulsars in Globular Clusters}

With the exception of those with detected gamma-ray pulsations,\footnote{Gamma-ray pulsations have been detected from a MSP in two globular clusters: PSR J1823-3021A located in the globular cluster NGC 6624~\cite{2011Sci...334.1107F,TheFermi-LAT:2013ssa}, and PSR B1821-24 within NGC 6626~\cite{Johnson:2013uza}.} the gamma-ray emission observed from an individual globular cluster provides us with little direct information about the pulsars within that system. In particular, it is not generally possible to determine whether the observed emission originates from a single luminous MSP, or from a larger number of fainter pulsars. Estimates for the number of gamma-ray emitting MSPs in a given globular cluster have sometimes been made by naively dividing the observed luminosity by a quantity taken to represent the gamma-ray luminosity of a typical MSP (for example, see Ref.~\cite{collaboration:2010bb}, which adopted $\langle  L_{\gamma }\rangle \simeq 1.4\times 10^{33}$ erg/s). For realistic pulsar luminosity functions, however, the total emission from a collection of such sources will often be dominated by a single luminous MSP, making such estimates quite unreliable.\footnote{Throughout this paper, gamma-ray luminosities refer to isotropic-equivalent values. The true luminosity of a given MSP, integrated over all angles, could be significantly higher or lower than this quantity.} 

Despite these challenges, one can in principle use an ensemble of globular clusters to make statistical inferences regarding the number of gamma-ray emitting MSPs they contain, and the gamma-ray luminosity function of those MSPs. In this section, we conduct such an analysis, utilizing the gamma-ray fluxes from globular clusters as reported in the previous section, and comparing these results to the stellar encounter rates (as previously calculated from observed kinematics~\cite{Bahramian:2013ihw}) and visible luminosities of those globular clusters.

\begin{figure}
\includegraphics[keepaspectratio,width=0.8\textwidth]{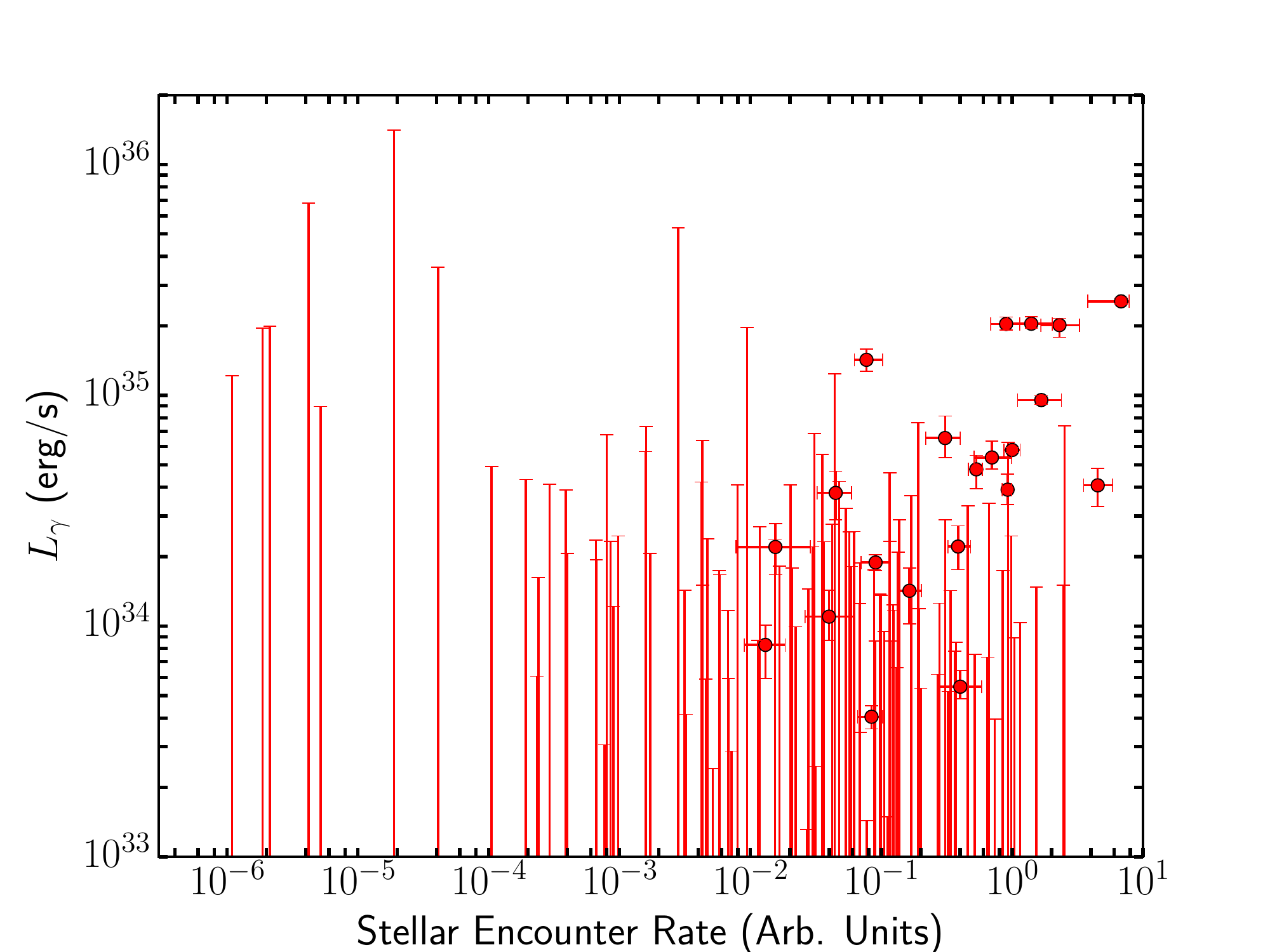}
\caption{The gamma-ray luminosity (0.1-100 GeV) of globular clusters as a function of their stellar encounter rate, as calculated in Ref.~\cite{Bahramian:2013ihw}. The stellar encounter rates are normalized to be equal to unity in the case of NGC 104 (47 Tuc).}
\label{GammaE}
\end{figure}

It generally is believed that most MSPs within globular clusters are generated through stellar interactions. If this is in fact the case, we should expect the average number of pulsars in such a system to be proportional to the cluster's stellar encounter rate. In our analysis, we make use of the stellar encounter rates as presented in Ref.~\cite{Bahramian:2013ihw}. In that study, the authors used the following expression to calculate the stellar encounter rate for a given globular cluster:
\begin{equation}
\Gamma_e = \frac{4 \pi}{\sigma_c} \int \rho^2(r) r^2 dr 
\end{equation}
where $\sigma_c$ is the velocity dispersion at the core radius of the cluster and $\rho(r)$ is the stellar density profile. In Fig.~\ref{GammaE}, we plot the stellar encounter rate verses the gamma-ray luminosity of the Milky Way's globular cluster population (for the 124 globular clusters with stellar encounter rates calculated in Ref~\cite{Bahramian:2013ihw}; see also Tables~\ref{table20}-\ref{table22}). As expected, those clusters with the highest stellar encounter rates also feature high gamma-ray luminosities, with some exceptions. Note that the stellar encounter rates are given in arbitrary units, normalized such that the value for NGC 104 (47 Tuc) is equal to unity.

\begin{table}
\renewcommand{\arraystretch}{1.2}
\begin{tabular}{|c|c|c|c|c|c|c|}
\hline 
Globular Cluster & Alt.~Name &  Distance~(kpc) & Luminosity~($L_{\odot}$)  & $\Gamma_e$ \tabularnewline
\hline 
\hline
 NGC 104     & 47 Tuc       &   4.46 & $5.01\times 10^{5} $ & $ 1.00_{-0.13}^{+0.15}$ \tabularnewline
 \hline
 NGC 362     &             &   8.61 & $2.01\times 10^{5} $ & $ 0.74_{-0.12}^{+0.14}$ \tabularnewline
 \hline
 NGC 1261    &              &  16.29 & $1.13\times 10^{5} $ & $ 0.015_{-0.004}^{+0.011 }$ \tabularnewline
 \hline
 Palomar 2       &              &  27.11 & $1.32\times 10^{5} $ & $ 0.93_{-0.56}^{+0.84 }$ \tabularnewline
 \hline
 NGC 1851    &               &  12.07 & $1.84\times 10^{5} $ & $ 1.53_{-0.19}^{+0.20 }$ \tabularnewline
 \hline
 NGC 1904    & M79        &  12.94 & $1.19\times 10^{5} $ & $ 0.12_{-0.014}^{+0.019 }$ \tabularnewline
 \hline
 NGC 2419    &              &  82.49 & $5.01\times 10^{5} $ & $ 0.0028^{+0.0008}_{-0.0005}$ \tabularnewline
 \hline
 NGC 2808    &              &   9.59 & $4.88\times 10^{5} $ & $ 0.92_{-0.083}^{+0.067 }$ \tabularnewline
 \hline
  NGC 4147    &               &  19.30 & $2.51\times 10^{4} $ & $ 0.017_{-0.006}^{+0.013}$ \tabularnewline
 \hline
 NGC 4372    &               &   5.81 & $1.12\times 10^{5} $ & $ 0.0002^{+0.0004}_{-0.0001}$ \tabularnewline
 \hline
 NGC 5024    & M53         &  17.85 & $2.61\times 10^{5} $ & $ 0.035_{-0.010}^{+0.012 }$ \tabularnewline
 \hline
 NGC 5139    & Omega Centauri     &   5.17 & $1.09\times 10^{6} $ & $ 0.090_{-0.020}^{+0.027 }$ \tabularnewline
 \hline
 NGC 5272    & M3          &  10.18 & $3.05\times 10^{5} $ & $ 0.19_{-0.018}^{+0.033 }$ \tabularnewline
 \hline
 NGC 5286    &              &  11.67 & $2.68\times 10^{5} $ & $ 0.46_{-0.061}^{+0.058 }$ \tabularnewline
 \hline
 NGC 5634    &               &  25.18 & $1.02\times 10^{5} $ & $ 0.020_{-0.008}^{+0.014 }$ \tabularnewline
 \hline
 NGC 5694    &              &  35.01 & $1.16\times 10^{5} $ & $ 0.19_{-0.034}^{+0.052 }$ \tabularnewline
 \hline
 NGC 5824    &              &  32.17 & $2.96\times 10^{5} $ & $ 0.98_{-0.16}^{+0.17 }$ \tabularnewline
 \hline
 NGC 5904    & M5          &   7.47 & $2.86\times 10^{5} $ & $ 0.16_{-0.030}^{+0.039 }$ \tabularnewline
 \hline
  NGC 5927    &             &   7.67 & $1.14\times 10^{5} $ & $ 0.068_{-0.010}^{+0.013 }$ \tabularnewline
 \hline
 NGC 5946    &               &  10.55 & $6.37\times 10^{4} $ & $ 0.13_{-0.045}^{+0.034 }$ \tabularnewline
 \hline
 NGC 5986    &               &  10.43 & $2.03\times 10^{5} $ & $ 0.062_{-0.010}^{+0.016 }$ \tabularnewline
 \hline
 NGC 6093    & M80         &  10.01 & $1.67\times 10^{5} $ & $ 0.53_{-0.069}^{+0.059 }$ \tabularnewline
 \hline
 NGC 6121    & M4           &   2.22 & $6.43\times 10^{4} $ & $ 0.027_{-0.010}^{+0.012 }$ \tabularnewline
 \hline
 NGC 6139    &               &  10.12 & $1.89\times 10^{5} $ & $ 0.31_{-0.089}^{+0.094 }$ \tabularnewline
 \hline
 NGC 6205    & M13        &   7.14 & $2.25\times 10^{5} $ & $ 0.069_{-0.015}^{+0.018 }$ \tabularnewline
 \hline
 NGC 6229    &              &  30.46 & $1.43\times 10^{5} $ & $ 0.048_{-0.009}^{+0.031 }$ \tabularnewline
 \hline
 NGC 6218    & M12         &   4.83 & $7.18\times 10^{4} $ & $ 0.013_{-0.0040}^{+0.0054}$ \tabularnewline
 \hline
 NGC 6254    & M10         &   4.39 & $8.39\times 10^{4} $ & $ 0.031_{-0.0041}^{+0.0043}$ \tabularnewline
 \hline
 NGC 6256    &              &  10.28 & $6.19\times 10^{4} $ & $ 0.17_{-0.060}^{+0.119 }$ \tabularnewline
 \hline
 NGC 6266    & M62          &   6.83 & $4.02\times 10^{5} $ & $ 1.67_{-0.57}^{+0.71 }$ \tabularnewline
 \hline
 NGC 6273    & M19         &   8.80 & $3.84\times 10^{5} $ & $ 0.20_{-0.039}^{+0.067 }$ \tabularnewline
 \hline
 NGC 6284    &               &  15.29 & $1.31\times 10^{5} $ & $ 0.67_{-0.11}^{+0.12 }$ \tabularnewline
 \hline
  \hline
\end{tabular}
\caption{The distance to each globular cluster, as well as its visible luminosity and stellar encounter rate ($\Gamma_e$), as calculated in Ref.~\cite{Bahramian:2013ihw}. The stellar encounter rates are normalized such that $\Gamma_e=1$
 for the case of NGC 104. We include in this table every globular cluster with either $\Gamma_e\ge 0.01$ or $L_{\rm V}\ge 10^5 \, L_{\odot}$.}
\label{table20}
\end{table}

\begin{table}
\renewcommand{\arraystretch}{1.2}
\begin{tabular}{|c|c|c|c|c|c|c|}
\hline 
Globular Cluster & Alt.~Name &  Distance~(kpc) & Luminosity~($L_{\odot}$)  & $\Gamma_e$ \tabularnewline
\hline 
 \hline
 NGC 6287    &              &   9.38 & $7.52\times 10^{4} $ & $ 0.036_{-0.0077}^{+0.0077 }$ \tabularnewline
 \hline
 NGC 6293    &               &   9.48 & $1.11\times 10^{5} $ & $ 0.85_{-0.24}^{+0.37 }$ \tabularnewline
 \hline
 NGC 6304    &              &   5.88 & $7.11\times 10^{4} $ & $ 0.12_{-0.022}^{+0.054 }$ \tabularnewline
 \hline
 NGC 6316    &              &  10.45 & $1.85\times 10^{5} $ & $ 0.077_{-0.015}^{+0.025 }$ \tabularnewline
 \hline
 NGC 6341    & M92          &   8.27 & $1.64\times 10^{5} $ & $ 0.27_{-0.029}^{+0.030 }$ \tabularnewline
 \hline
 NGC 6325    &               &   7.83 & $5.20\times 10^{4} $ & $ 0.12_{-0.046}^{+0.045 }$ \tabularnewline
 \hline
 NGC 6333    & M9           &   7.91 & $1.29\times 10^{5} $ & $ 0.13_{-0.042}^{+0.059 }$ \tabularnewline
 \hline
 NGC 6342    &               &   8.53 & $3.16\times 10^{4} $ & $ 0.045_{-0.013}^{+0.014 }$ \tabularnewline
 \hline
 NGC 6356    &               &  15.08 & $2.17\times 10^{5} $ & $ 0.088_{-0.014}^{+0.020 }$ \tabularnewline
 \hline
 NGC 6355    &               &   9.22 & $1.45\times 10^{5} $ & $ 0.10_{-0.026}^{+0.041 }$ \tabularnewline
 \hline
 Terzan 2    & HP 3          &   7.49 & $1.92\times 10^{4} $ & $ 0.022_{-0.014}^{+0.029 }$ \tabularnewline
 \hline
 NGC 6380    & Ton 1        &  10.88 & $8.55\times 10^{4} $ & $ 0.12_{-0.045}^{+0.068 }$ \tabularnewline
 \hline
 NGC 6388    &               &   9.92 & $4.97\times 10^{5} $ & $ 0.90_{-0.21}^{+0.24 }$ \tabularnewline
 \hline
 NGC 6402    & M14         &   9.25 & $3.73\times 10^{5} $ & $ 0.12_{-0.030}^{+0.032 }$ \tabularnewline
 \hline
 NGC 6401    &               &  10.56 & $1.24\times 10^{5} $ & $ 0.044_{-0.011}^{+0.011 }$ \tabularnewline
 \hline
 NGC 6397    &               &   2.30 & $3.87\times 10^{4} $ & $ 0.084_{-0.018}^{+0.018 }$ \tabularnewline
 \hline
 Palomar 6       &               &   5.79 & $4.45\times 10^{4} $ & $ 0.016_{-0.008}^{+0.013 }$ \tabularnewline
 \hline
 Terzan 5    & Terzan 11     &   5.98 & $7.94\times 10^{4} $ & $ 6.80_{-3.02}^{+1.04 }$ \tabularnewline
 \hline
 NGC 6440    &               &   8.45 & $2.70\times 10^{5} $ & $ 1.40_{-0.48}^{+0.63 }$ \tabularnewline
 \hline
 NGC 6441    &               &  11.60 & $6.08\times 10^{5} $ & $ 2.30_{-0.64}^{+0.97 }$ \tabularnewline
 \hline
 Terzan 6    & HP 5        &   6.78 & $9.29\times 10^{4} $ & $ 2.47_{-1.72}^{+5.07 }$ \tabularnewline
 \hline
 NGC 6453    &              &  11.57 & $6.61\times 10^{4} $ & $ 0.37_{-0.09}^{+0.13 }$ \tabularnewline
 \hline
 NGC 6517    &              &  10.63 & $1.71\times 10^{5} $ & $ 0.34_{-0.10}^{+0.15 }$ \tabularnewline
 \hline
 NGC 6522    &              &   7.70 & $9.82\times 10^{4} $ & $ 0.36_{-0.10}^{+0.11 }$ \tabularnewline
 \hline
 NGC 6528    &              &   7.93 & $3.63\times 10^{4} $ & $ 0.28_{-0.05}^{+0.11 }$ \tabularnewline
 \hline
 NGC 6539    &               &   7.79 & $1.77\times 10^{5} $ & $ 0.042_{-0.015}^{+0.029 }$ \tabularnewline
  \hline
 NGC 6544    &              &   2.96 & $5.11\times 10^{4} $ & $ 0.11_{-0.037}^{+0.068 }$ \tabularnewline
 \hline
 NGC 6541    &               &   7.54 & $2.19\times 10^{5} $ & $ 0.39_{-0.063}^{+0.095 }$ \tabularnewline
 \hline
 NGC 6553    &               &   5.96 & $1.10\times 10^{5} $ & $ 0.069_{-0.019}^{+0.027 }$ \tabularnewline
 \hline
 NGC 6558    &              &   7.37 & $3.22\times 10^{4} $ & $ 0.11_{-0.019}^{+0.026 }$ \tabularnewline
\hline 
 NGC 6569    &               &  10.90 & $1.75\times 10^{5} $ & $ 0.054_{-0.021}^{+0.030 }$ \tabularnewline
 \hline
 NGC 6584    &               &  13.49 & $1.02\times 10^{5} $ & $ 0.012_{-0.0034}^{+0.0054 }$ \tabularnewline 
\hline
   \hline
\end{tabular}
\caption{Continued from Table~\ref{table20}. The distance to each globular cluster, as well as its visible luminosity and stellar encounter rate ($\Gamma_e$), as calculated in Ref.~\cite{Bahramian:2013ihw}. The stellar encounter rates are normalized such that $\Gamma_e=1$
 for the case of NGC 104. We include in this table every globular cluster with either $\Gamma_e\ge 0.01$ or $L_{\rm V}\ge 10^5 \, L_{\odot}$.}
\label{table21}
\end{table}

\begin{table}
\renewcommand{\arraystretch}{1.2}
\begin{tabular}{|c|c|c|c|c|c|c|}
\hline 
Globular Cluster & Alt.~Name &  Distance~(kpc) & Luminosity~($L_{\odot}$)  & $\Gamma_e$ \tabularnewline
\hline 
 \hline
  NGC 6624    &              &   7.91 & $8.47\times 10^{4} $ & $ 1.15_{-0.18}^{+0.11 }$ \tabularnewline
 \hline
 NGC 6626    & M28          &   5.52 & $1.57\times 10^{5} $ & $ 0.65_{-0.091}^{+0.084 }$ \tabularnewline
 \hline
 NGC 6638    &               &   9.41 & $6.03\times 10^{4} $ & $ 0.14_{-0.027}^{+0.039 }$ \tabularnewline
 \hline
 NGC 6637    & M69          &   8.80 & $9.73\times 10^{4} $ & $ 0.090_{-0.018}^{+0.036 }$ \tabularnewline
 \hline
 NGC 6642    &               &   8.13 & $3.94\times 10^{4} $ & $ 0.10_{-0.025}^{+0.031 }$ \tabularnewline
 \hline
 NGC 6652    &               &  10.00 & $3.94\times 10^{4} $ & $ 0.70_{-0.19}^{+0.29 }$ \tabularnewline
 \hline
 NGC 6656    & M22         &   3.23 & $2.15\times 10^{5} $ & $ 0.078_{-0.026}^{+0.032 }$ \tabularnewline
 \hline
 NGC 6681    & M 70          &   9.01 & $6.03\times 10^{4} $ & $ 1.04_{-0.19}^{+0.27 }$ \tabularnewline
 \hline
 NGC 6712    &              &   6.93 & $8.55\times 10^{4} $ & $ 0.031_{-0.0066}^{+0.0056 }$ \tabularnewline
 \hline
 NGC 6715    & M54          &  26.49 & $8.39\times 10^{5} $ & $ 2.52_{-0.27}^{+0.23 }$ \tabularnewline
 \hline
 NGC 6717    & Palomar 9        &   7.11 & $1.57\times 10^{4} $ & $ 0.040_{-0.014}^{+0.022 }$ \tabularnewline
 \hline 
 NGC 6723    &            &   8.65 & $1.16\times 10^{5} $ & $ 0.011_{-0.0044}^{+0.0080 }$ \tabularnewline
 \hline
 NGC 6752    &               &   3.99 & $1.06\times 10^{5} $ & $ 0.40_{-0.13}^{+0.18 }$ \tabularnewline
 \hline
 NGC 6760    &               &   7.36 & $1.17\times 10^{5} $ & $ 0.057_{-0.019}^{+0.027 }$ \tabularnewline
 \hline
 NGC 6779    & M 56         &   9.44 & $7.87\times 10^{4} $ & $ 0.028_{-0.009}^{+0.012 }$ \tabularnewline
\hline
 Palomar 10      &             &   5.93 & $1.77\times 10^{4} $ & $ 0.059_{-0.036}^{+0.043 }$ \tabularnewline
 \hline
 Palomar 11      &               &  13.40 & $5.01\times 10^{4} $ & $ 0.021_{-0.007}^{+0.011 }$ \tabularnewline
 \hline
 NGC 6864    & M 75          &  20.84 & $2.29\times 10^{5} $ & $ 0.31_{-0.082}^{+0.095 }$ \tabularnewline
 \hline
 NGC 6934    &               &  15.63 & $8.17\times 10^{4} $ & $ 0.030_{-0.008}^{+0.012 }$ \tabularnewline
 \hline
 NGC 7006    &               &  41.21 & $1.00\times 10^{5} $ & $ 0.0094^{+0.0049}_{-0.0033}$ \tabularnewline
 \hline
 NGC 7078    & M 15         &  10.38 & $4.06\times 10^{5} $ & $ 4.51_{-0.99}^{+1.36 }$ \tabularnewline
 \hline
 NGC 7089    & M 2          &  11.56 & $3.50\times 10^{5} $ & $ 0.52_{-0.071}^{+0.078 }$ \tabularnewline
 \hline
 NGC 7099    & M 30          &   8.12 & $8.17\times 10^{4} $ & $ 0.32_{-0.08}^{+0.12 }$ \tabularnewline
 \hline
\hline 
\end{tabular}
\caption{Continued from Tables~\ref{table20}-\ref{table21}. The distance to each globular cluster, as well as its visible luminosity and stellar encounter rate ($\Gamma_e$), as calculated in Ref.~\cite{Bahramian:2013ihw}. The stellar encounter rates are normalized such that $\Gamma_e=1$
 for the case of NGC 104. We include in this table every globular cluster with either $\Gamma_e\ge 0.01$ or $L_{\rm V}\ge 10^5 \, L_{\odot}$.}
\label{table22}
\end{table}

\begin{figure}
\includegraphics[keepaspectratio,width=0.8\textwidth]{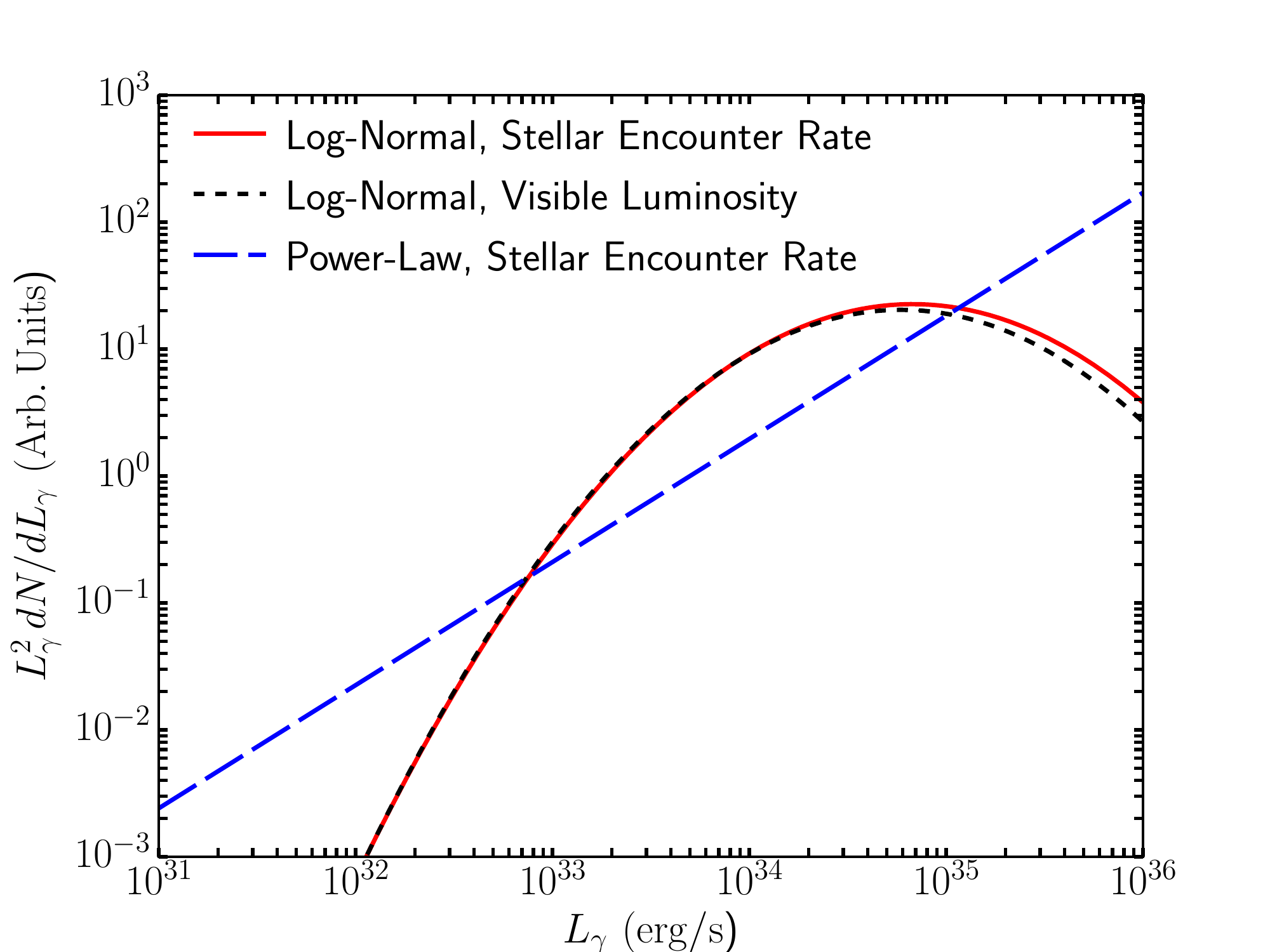}
\caption{The gamma-ray luminosity function of millisecond pulsars in globular clusters, for the best-fit parameter values found by our fit. Results are shown assuming that the average number of pulsars is proportional to the stellar encounter rate, or to the visible luminosity of the cluster. We also show results using a log-normal or power-law form for the luminosity function.}
\label{lumfunc}
\end{figure}

\begin{figure}
\includegraphics[keepaspectratio,width=0.8\textwidth]{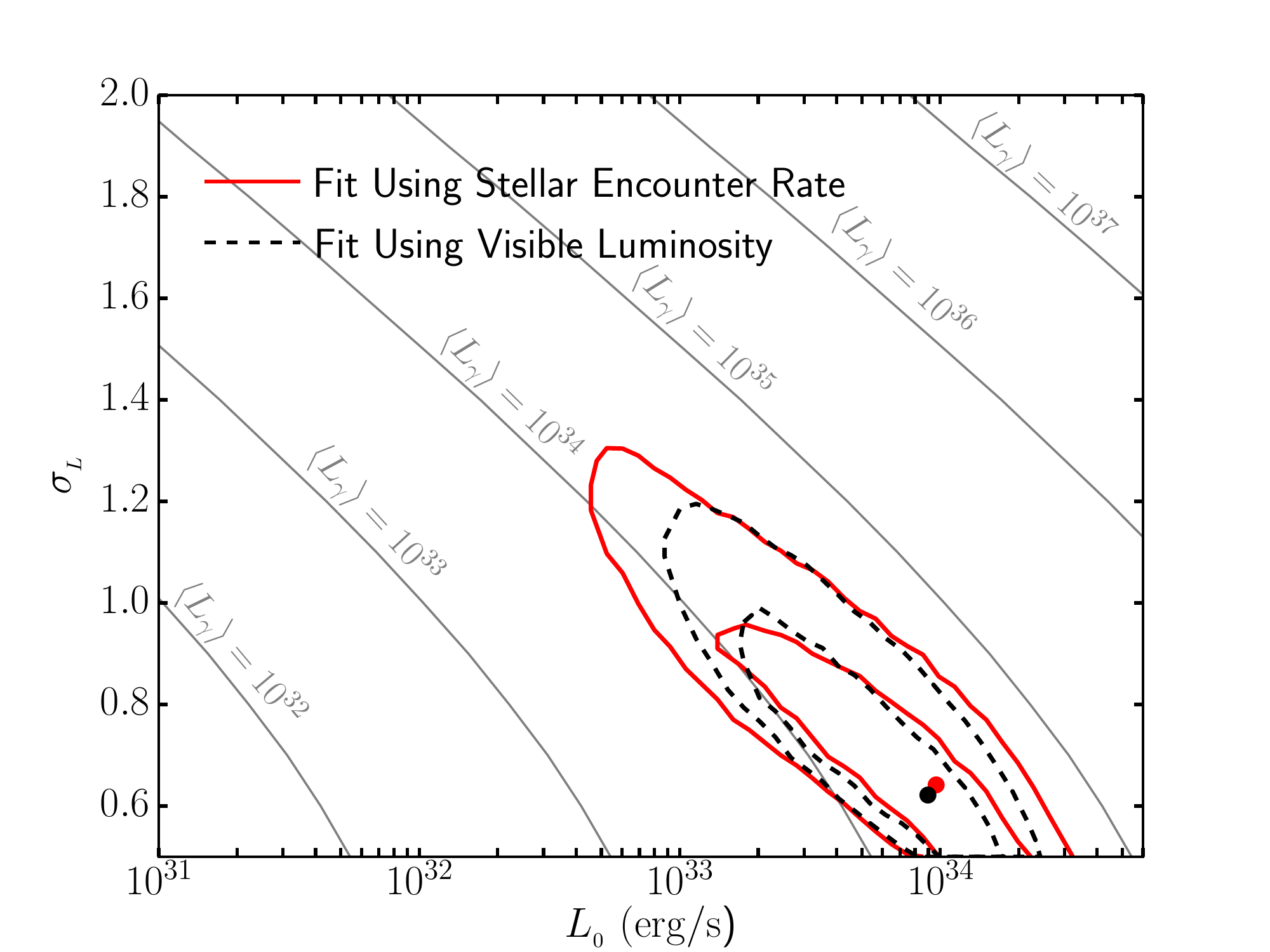}
\caption{Constraints (1$\sigma$, 2$\sigma$) on the gamma-ray luminosity function parameters found by our fit. Results are shown assuming that the average number of pulsars is proportional to the stellar encounter rate (red solid), or to the visible luminosity of the cluster (black dashed). The grey lines represent contours of constant mean pulsar luminosity. Our fit consistently prefers luminosity functions with a mean luminosity in the range of $\sim$$(0.7-5)\times 10^{34}$ erg/s.}
\label{lzerosigma1}
\end{figure}

Throughout most of this study, we parameterize the gamma-ray luminosity function of MSPs as follows, using a (base 10) log-normal distribution:
\begin{eqnarray}
\frac{dN}{d\log_{10} L_{\gamma}} \equiv \frac{1}{\sigma_{_L} \sqrt{2\pi}} \, \exp\bigg[-\frac{(\log_{10} L_{\gamma} - \log_{10} L_{_0})^2}{2\sigma_{_L}^2}\bigg],
\end{eqnarray}
or equivalently
\begin{eqnarray}
\frac{dN}{dL_{\gamma}} \equiv \frac{\log_{10} e}{\sigma_{_L} \sqrt{2\pi} L_{\gamma}} \, \exp\bigg[-\frac{(\log_{10} L_{\gamma} - \log_{10} L_{_0})^2}{2\sigma_{_L}^2}\bigg].
\end{eqnarray}
In these expressions, we treat $L_0$ and $\sigma_L$ as free parameters, which we will constrain in our analysis. We also treat as a free parameter the average number of MSPs per unit stellar encounter rate, $R_{\rm MSP}$.

To constrain these parameters, we have utilized a Monte Carlo simulation. For each value of the stellar encounter rate, we draw from a Poisson distribution to determine the number of MSPs in a given globular cluster, and then draw from the luminosity function for each of those MSPs to determine the total gamma-ray luminosity of the cluster. After performing this exercise for $10^6$ globular clusters with a given stellar encounter rate, we arrive at a probability distribution for the gamma-ray luminosity of a given cluster (for each combination of $L_0$, $\sigma_L$ and $R_{\rm MSP}$). We then compare this distribution to the data, as shown in Fig.~\ref{GammaE}, to determine the range of these parameters that provides a good fit.

Overall, our fit favors the following range of parameter values:  $L_0=\big[ 0.88^{+0.79}_{-0.41} \big] \times 10^{34}$ erg/s, $\sigma_L=0.62^{+0.15}_{-0.16}$, and $R_{\rm MSP} =1.79^{+0.77}_{-0.50}$, where the errors represent the 1$\sigma$ uncertainty after marginalizing over the values of the other two parameters. In Fig.~\ref{lumfunc}, the solid red line represents the luminosity function corresponding to the best-fit values of these parameters. In Fig.~\ref{lzerosigma1}, we present the regions of the parameter space (in terms of the $L_0$-$\sigma_L$ plane) that are favored by our fit at the 1$\sigma$ and 2$\sigma$ confidence levels.

\begin{figure}
\includegraphics[keepaspectratio,width=0.4950\textwidth]{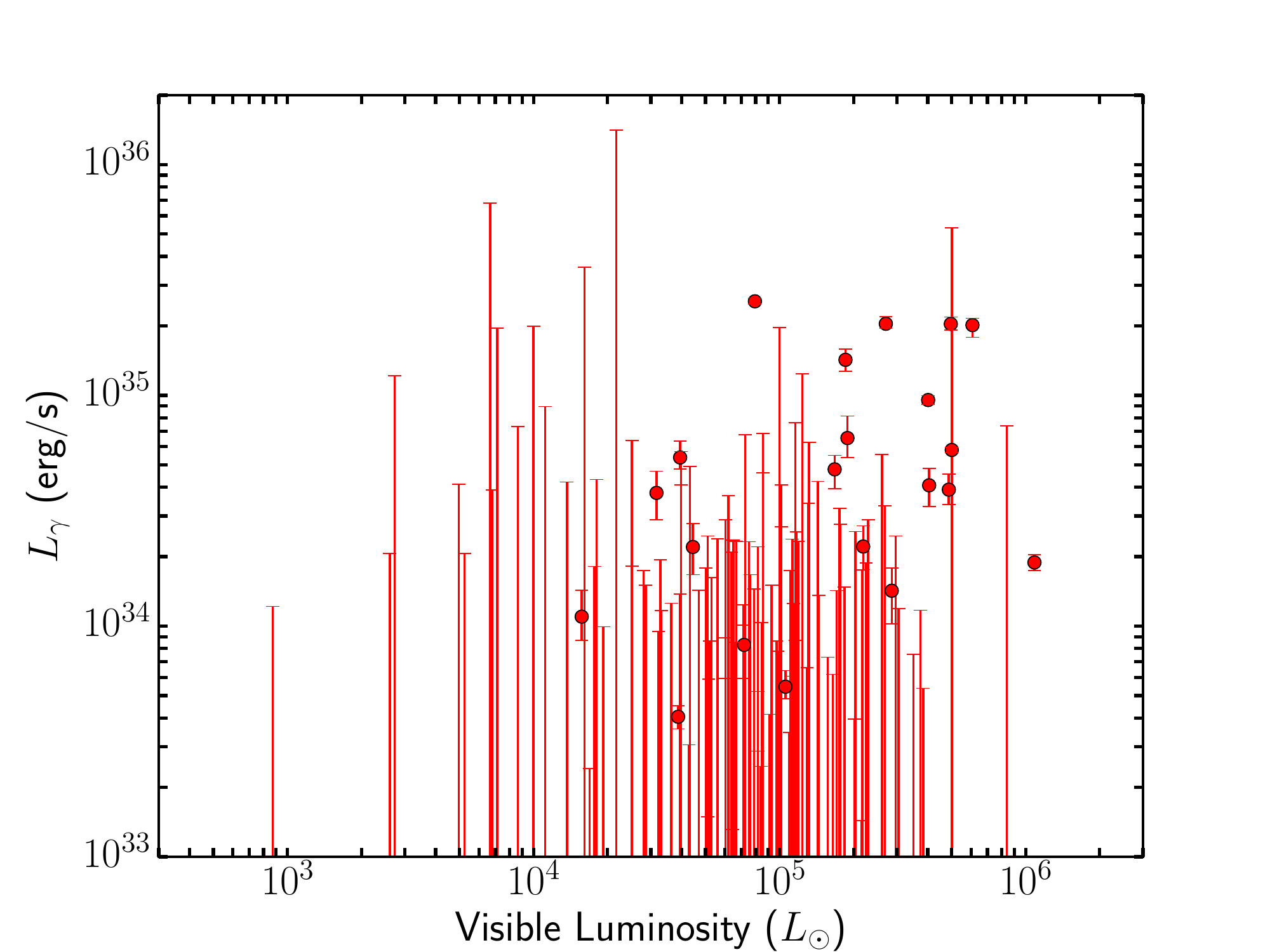}
\includegraphics[keepaspectratio,width=0.4950\textwidth]{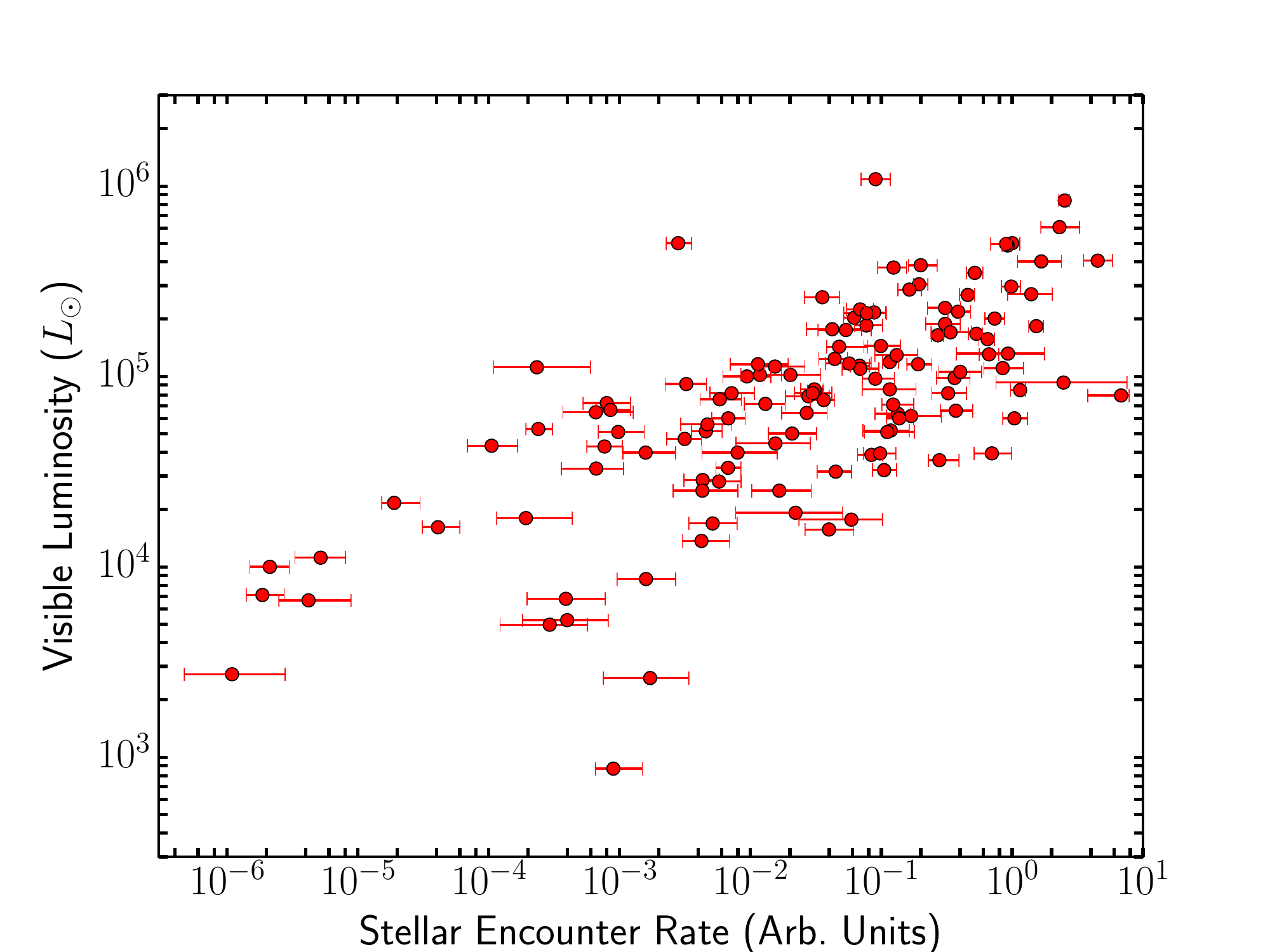}
\caption{In the left frame, we show a comparison similar to Fig.~\ref{GammaE}, but using the visible luminosity of each globular cluster (rather than the stellar encounter rate). In the right frame, we compare the visible luminosity to the stellar encounter rate of each globular cluster, finding a high degree of correlation.}
\label{Vlum}
\end{figure}

Although it is generally assumed that the MSP population in globular clusters is generated through stellar interactions, it is not known with complete certainty that this is the case. With this in mind, we have repeated the procedure described above, but instead assuming that the average number of MSPs in a given cluster is proportional to the visible luminosity of the cluster (corresponding roughly to the total number of stars). In the left frame of Fig.~\ref{Vlum}, we compare the visible luminosity of each globular cluster to its gamma-ray luminosity (see Tables~\ref{table20}-\ref{table22}). We find that the clusters with the highest visible luminosities also generally exhibit high gamma-ray luminosities. In the right frame of this figure, we compare the visible luminosities of these clusters to their stellar encounter rates. The correlation between these two quantities causes our Monte Carlo to yield similar results, regardless of whether we assume that the average number of pulsars is proportional to the stellar encounter rate or to the visible luminosity of the cluster. In the later case, we find a best-fit for $L_0=0.82 \times 10^{34}$ erg/s, $\sigma_L=0.61$, and $1.4 \times10^{-6}$ MSPs per solar luminosity. Results for this case are shown as black dashed lines in Figs.~\ref{lumfunc} and~\ref{lzerosigma1}.

\begin{figure}
\includegraphics[keepaspectratio,width=0.49\textwidth]{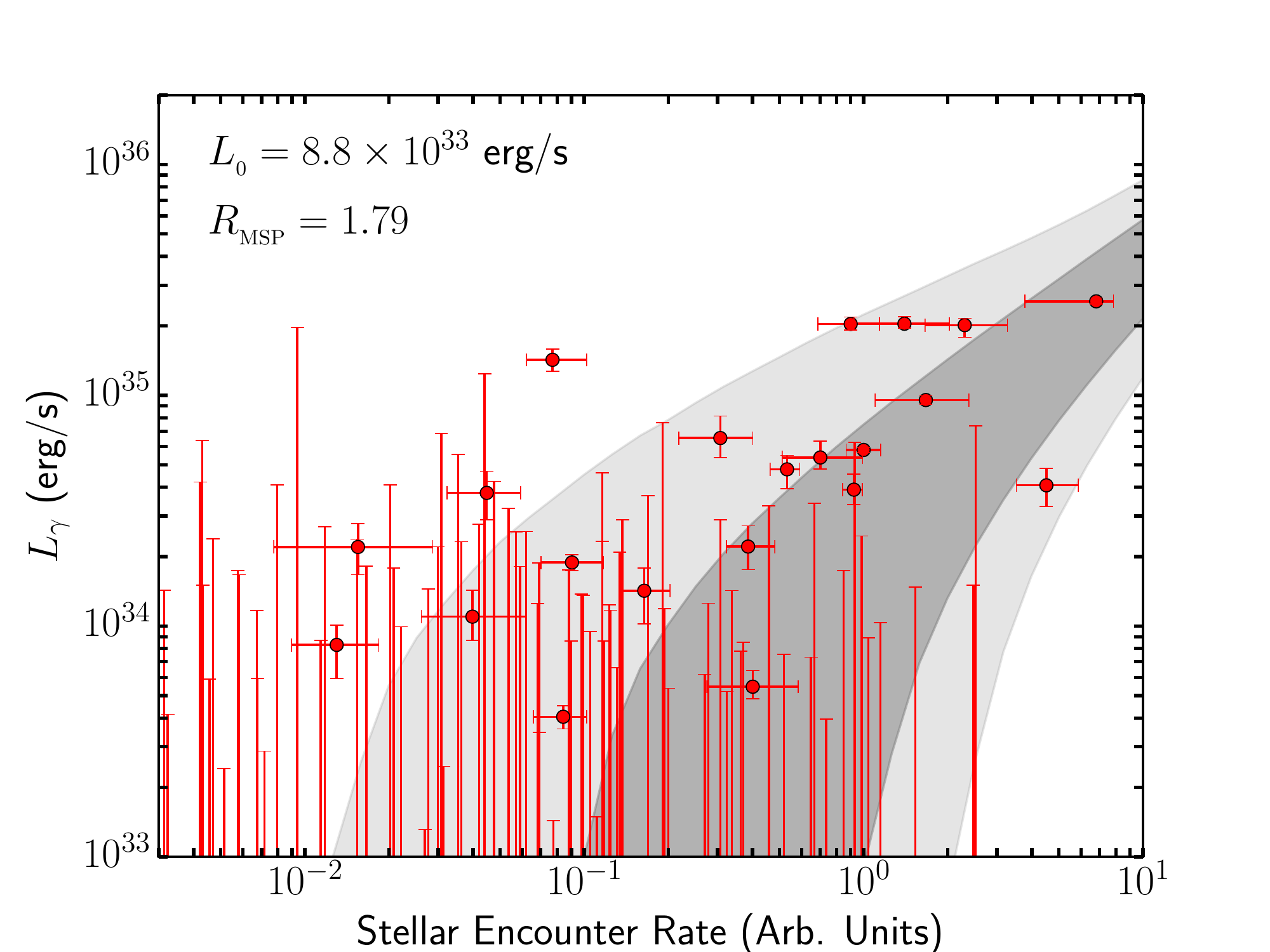}
\includegraphics[keepaspectratio,width=0.49\textwidth]{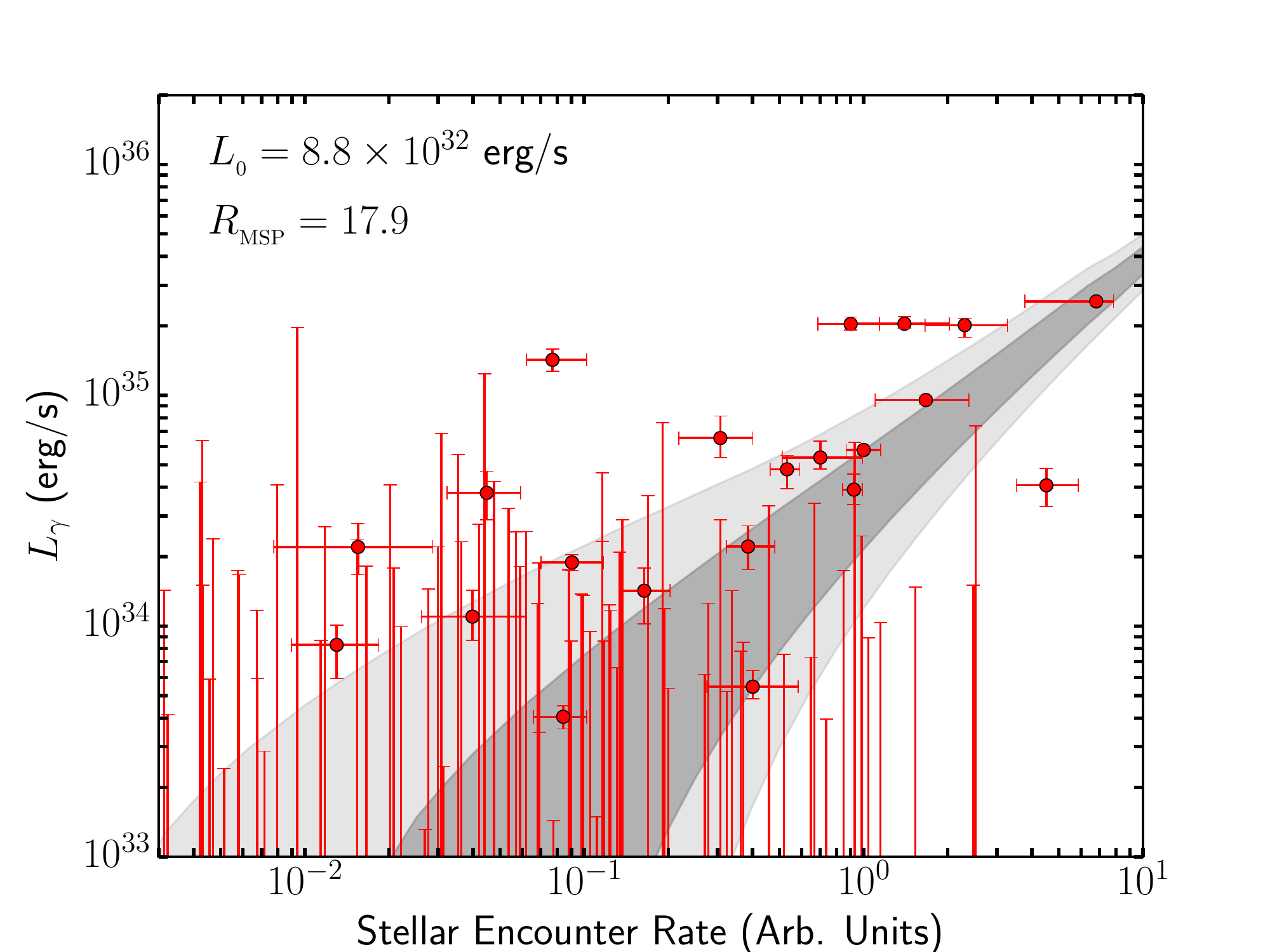}
\caption{The gamma-ray luminosity of globular clusters as a function of their stellar encounter rate, compared to the distributions predicted using two different luminosity functions. In the left frame, we adopt our best-fit parameters ($L_0= 8.8 \times 10^{33}$ erg/s, $\sigma_L=0.62$, and $R_{\rm MSP} =1.79$), while in the right frame we consider a model with ten times as many pulsars ($R_{\rm MSP} =17.9$), with ten times lower luminosity ($L_0= 8.8 \times 10^{32}$ erg/s). The shaded bands in each frame denote the regions in which the central 68.3\% (dark grey) and 95.5\% (light grey) of globular clusters are predicted to reside. Each of these cases predict the same mean luminosity ($3.9\times 10^{34}$ erg/s per unit stellar encounter rate), but very different distributions and median values.}
\label{bands}
\end{figure}

The discriminating power of our analysis is in large part driven by the width of the distribution of gamma-ray luminosities around the mean value. This is illustrated in Fig.~\ref{bands}, where we plot the regions in which the central 68.3\% (dark grey) and 95.5\% (light grey) of globular clusters are predicted to reside (for a given stellar encounter rate). In the left frame of this figure, we calculate these bands using our best-fit luminosity function parameters ($L_0= 8.8 \times 10^{33}$ erg/s, $\sigma_L=0.62$, and $R_{\rm MSP} =1.79$), while in the right frame we adopt values of $L_0$ and $R_{\rm MSP}$ that are ten times smaller and larger, respectively ($L_0= 8.8 \times 10^{32}$ erg/s, $\sigma_L=0.62$, and $R_{\rm MSP} =17.9$). While each of these cases predicts the same mean cluster luminosity (for a given stellar encounter rate), the widths of these distributions are very different. In particular, while our best-fit parameters provide a good fit to the observed distribution, the parameters adopted in the right frame clearly cannot account for the observed outliers.


\begin{figure}
\includegraphics[keepaspectratio,width=0.7\textwidth]{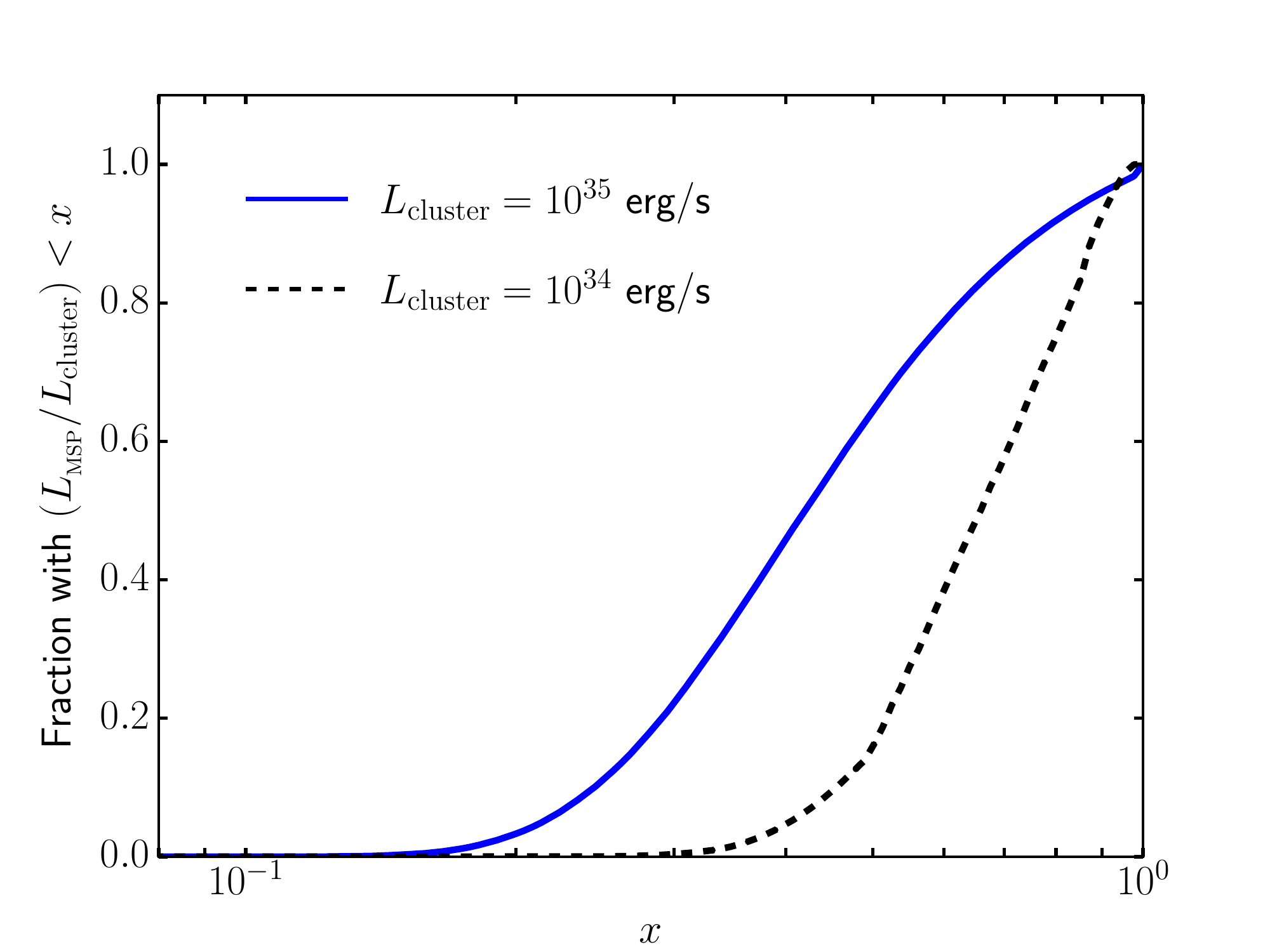}
\caption{For our best-fit luminosity function parameters ($L_0= 8.8 \times 10^{33}$ erg/s, $\sigma_L=0.62$), we plot the fraction of  globular clusters for which less than a given fraction of their total gamma-ray luminosity ($x$) comes from a single MSP ($L_{\rm MSP}$ denotes the luminosity of the brightest pulsar in the cluster). For approximately 85\% (37\%) of globular clusters with a total gamma-ray luminosity of $10^{34}$ erg/s ($10^{35}$ erg/s), more than half of their total gamma-ray emission originates from a single MSP.}
\label{fracprob}
\end{figure}

One important consequence of the MSP gamma-ray luminosity function derived here is that it predicts that many or most of the globular clusters detected by Fermi will be dominated (in terms of gamma-ray luminosity) by only one or two MSPs. More specifically, from the results shown in Fig.~\ref{fracprob}, we find that a single MSP provides more than half of the total gamma-ray luminosity in 85\% (37\%) of globular clusters with a total gamma-ray luminosity of $10^{34}$ erg/s ($10^{35}$ erg/s).

In order to test the impact of our assumption that the gamma-ray luminosities of MSPs follow a log-normal distribution, we alternatively consider a power-law form of the MSP luminosity function. Performing our analysis again (assuming an average number of MSPs that is proportional to the stellar encounter rate), but this time assuming a luminosity function that is given by $dN/dL_{\gamma} \propto L_{\gamma}^{\beta}$, over the range $10^{31} < L_{\gamma} < 10^{36}$ erg/s, we find that the best fit occurs for $\beta \simeq -1.03 \pm 0.06$ and $R_{\rm MSP}=1.18^{+0.42}_{-0.16}$. This provides a fit that is approximately as good as that found using a log-normal luminosity function. The best-fit power-law luminosity function is shown as a blue long-dashed line in Fig.~\ref{lumfunc}. In this case, the best-fit parameters yield a mean MSP luminosity of $\langle L_{\gamma} \rangle \simeq 7.5\times 10^{34}$ erg/s, which is slightly higher that than found in the log-normal case.


In principle, we can use the value of $R_{\rm MSP}$ derived in our analysis, combined with the sum of the stellar encounter rates (or visible luminosities), to constrain the total number of MSPs that are present within the Milky Way's system of globular clusters.  For our best-fit log-normal (power-law) parameters, derived assuming a pulsar population proportional to the the stellar encounter rate, this exercise yields an estimate that there are a total of $N_{\rm MSP}=76.1^{+32.7}_{-21.3}$ ($50.2^{+17.8}_{-6.8}$) MSPs that reside within this collection of 124 globular clusters. If we instead assume that the number of pulsars scales with the visible luminosity of a given cluster, we arrive at a slightly higher best-fit value of 87.4 MSPs. 

Taken at face value, this population of MSPs may seem surprisingly small. In particular, radio emission has already been detected from 129 MSPs residing in 25 globular clusters.\footnote{For a list of radio MSPs in globular clusters, see: \url{http://www.naic.edu/~pfreire/GCpsr.html}.} It is important to appreciate, however, that the analysis presented here is not very sensitive to the number of low-luminsity MSPs (those with $L_{\gamma} \lsim 3 \times 10^{33}$ erg/s). Although our method is able to constrain the fraction of the total gamma-ray emission that comes from such objects, it tells us little about the total number of such pulsars present. As a result, the total number of low-luminosity pulsars derived in our analysis is largely fixed by the parameterization we have adopted for the luminosity function, and we do not meaningfully constrain the number of MSPs that are present in the low-luminosity tail of the luminosity function.

The gamma-ray luminosity of a given MSP is given by:
\begin{eqnarray}
\label{lum}
L_{\gamma} &=& \eta_{\gamma} \, \dot{E} \\
&=& \eta_{\gamma} \, \frac{4 \pi^2 I \dot{P}}{P^3} \nonumber \\
&\simeq & 9.6 \times 10^{33} \, {\rm erg/s} \,\, \bigg(\frac{\eta_{\gamma}}{0.2}\bigg) \,  \bigg(\frac{B}{10^{8.5}\, {\rm G}}\bigg)^2 \, \bigg(\frac{3 \, {\rm ms}}{P}\bigg)^4, \nonumber
\end{eqnarray}
where $\eta_{\gamma}$ is the gamma-ray efficiency, $\dot{E}$ is the spin-down rate, $I$ is the neutron star's moment of inertia (taken to be $10^{45}$ g cm$^2$) and $P$ and $\dot{P}$ the rotational period of the pulsar and its time derivative. In the last step, we have used the following expression for $\dot{P}$ resulting from magnetic-dipole breaking:
\begin{equation}
\dot{P} \simeq 3.3 \times 10^{-20} \, \bigg(\frac{B}{10^{8.5}\, {\rm G}}\bigg)^2 \, \bigg(\frac{3 \, {\rm ms}}{P}\bigg),
\label{pdot}
\end{equation}
where $B$ is the strength of the magnetic field. 

Although a pulsar's gamma-ray luminosity is determined by its values of $P$, $B$ and $\eta_{\gamma}$, the quantities $B$ and $\eta_{\gamma}$ are in general not directly measureable. For pulsars observed in the field of the Milky Way, however, the values $P$ and $\dot{P}$ can be measured independently of $L_{\gamma}$, making it possible to infer the distributions of the underlying quantities. For pulsars in globular clusters, however, observed values of $\dot{P}$ are significantly impacted by accelerating forces acting on the neutron star, making it difficult to determine the intrinsic spin-down rate~\cite{Freire:2001qr}. In contrast, the periods of many MSPs in globular clusters have been measured at radio wavelengths.

\begin{figure}
\includegraphics[keepaspectratio,width=0.8\textwidth]{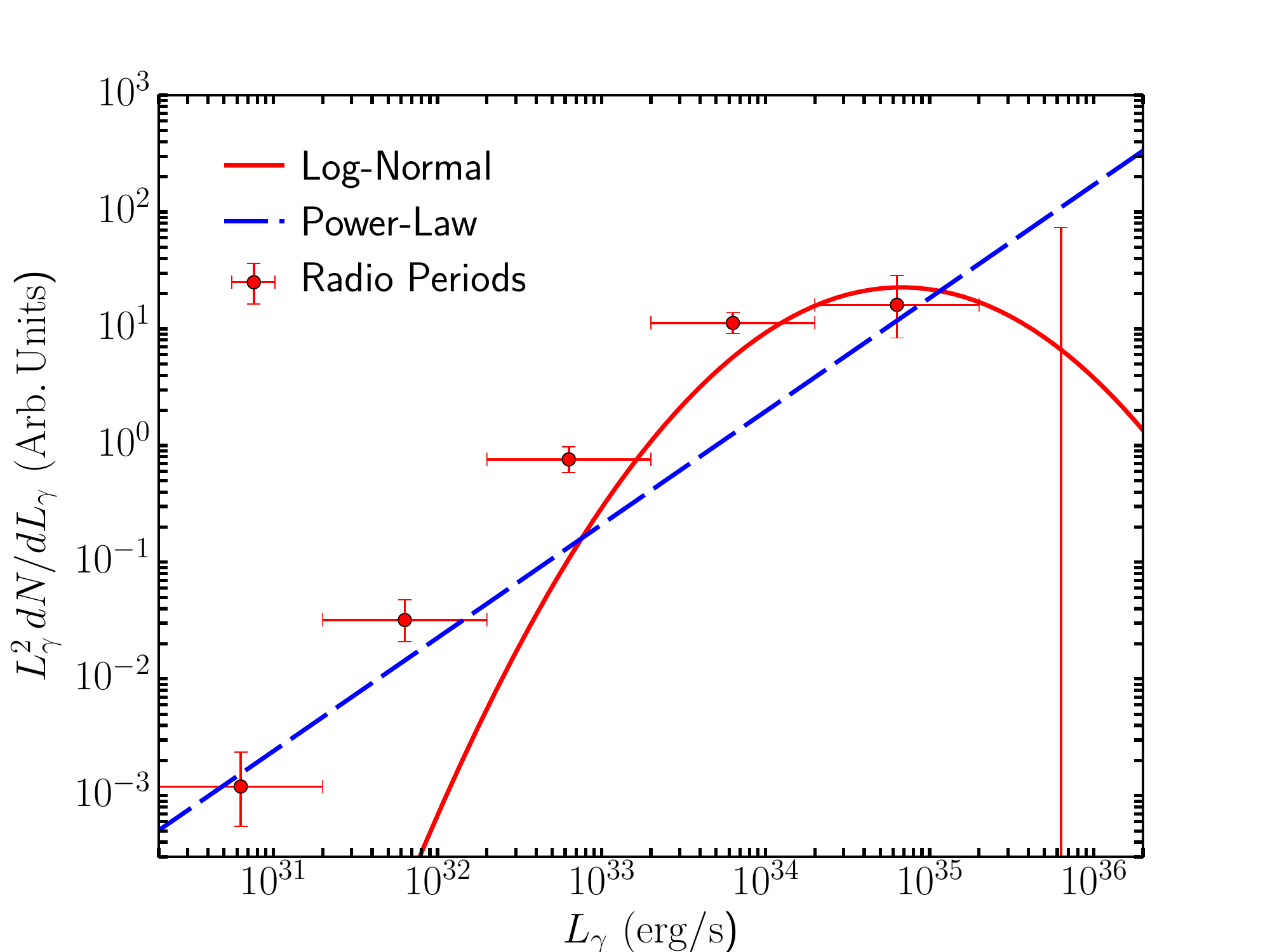}
\caption{The error bars depict the gamma-ray luminosity function of MSPs in globular clusters, derived from the observed distribution of radio periods, naively assuming $B=10^{8.5}$ G and $\eta_{\gamma}=0.2$ for each pulsar, making $L_{\gamma}$ entirely dependent on the pulsar's period. Although this result is in reasonable agreement with those luminosity functions found in this study (red solid and blue dashed lines) for $L_{\gamma} \gsim 10^{33}$ erg/s, the radio periods suggest that the true luminosity function has a significantly larger low-luminosity tail (which our method is not sensitive to).}
\label{radio}
\end{figure}

\begin{figure}
\includegraphics[keepaspectratio,width=0.8\textwidth]{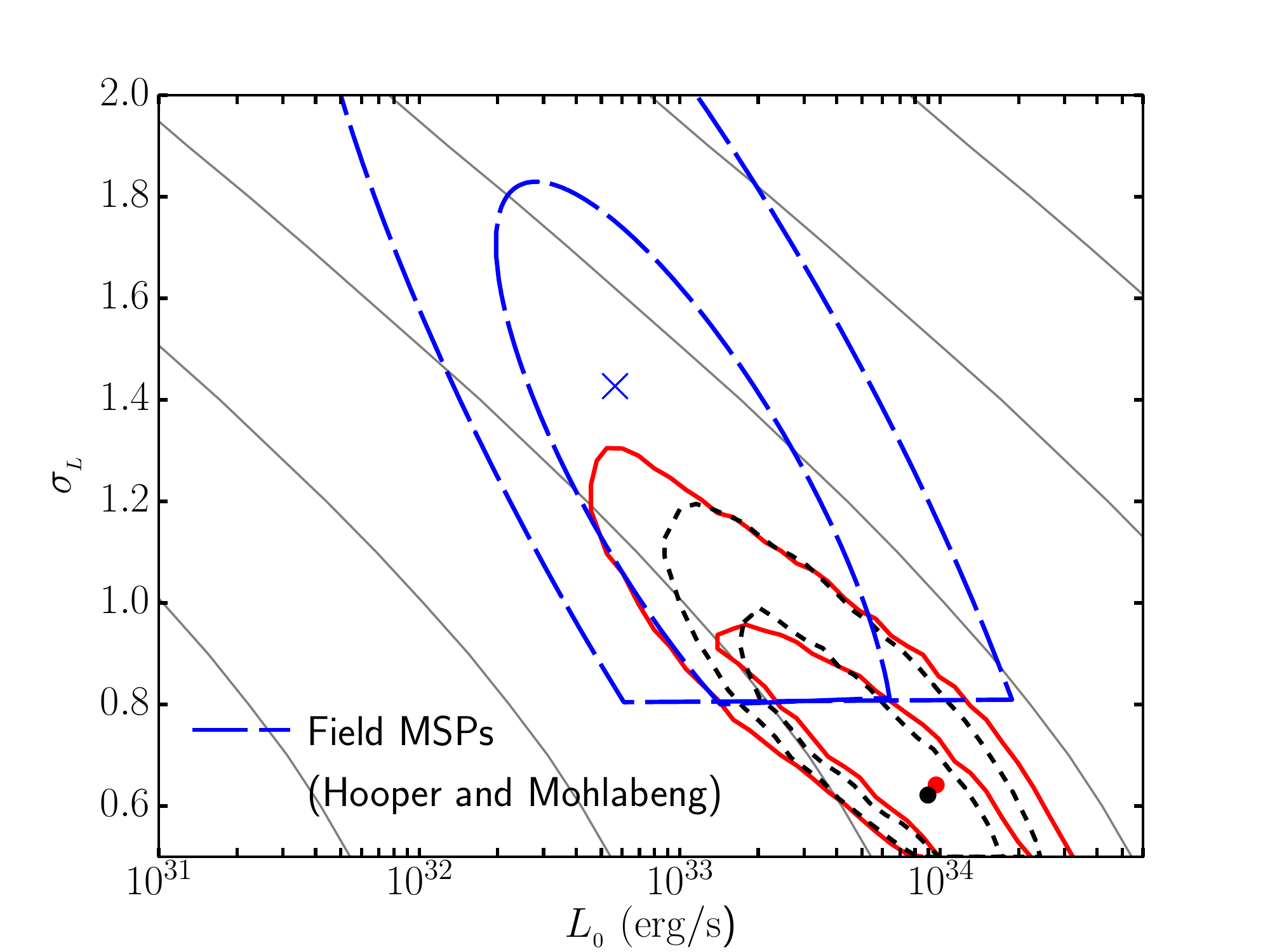}
\caption{The luminosity function parameters favored by this paper's analysis (solid red and dashed black contours, as in Fig.~\ref{lzerosigma1}) and as found previously for those MSPs {\it not} residing in globular clusters (long dashed blue) ~\cite{Hooper:2015jlu}. As these results overlap at the 1$\sigma$ level, it suggests that the gamma-ray luminosity function of these two pulsar populations may be quite similar. Note that Ref.~\cite{Hooper:2015jlu} adopted a $\sigma_{L} \ge 0.8$ prior, explaining the edge in the region shown. The grey lines denote contours of constant mean luminosity, with values as labeled in Fig.~\ref{lzerosigma1}.}
\label{lzerosigma}
\end{figure}

If we naively (and unrealistically) imagine that all MSPs have the same values of $B$ and $\eta_{\gamma}$, we can derive the gamma-ray luminosity function from the observed distribution of rotational periods. Given that the gamma-ray luminosity of an MSP scales with the fourth power of its inverse period, it is not entirely unreasonable to expect variations in $P$ to dominate the overall luminosity function. We show the result of this exercise in Fig.~\ref{radio}. For reasonable parameter values ($B=10^{8.5}$ G, $\eta_{\gamma}=0.2$), the resulting luminosity function is in good agreement with that derived in this study for $L_{\gamma} \gsim 10^{33}$ erg/s. At lower luminosities, however, the radio period distribution suggests that there are likely significantly more MSPs than are predicted by the luminosity function found in this study. As stated before, our method is largely insensitive to the number of pulsars with $L_{\gamma} \lsim 3 \times 10^{33}$ erg/s, so we should not be surprised by this result. If the true luminosity function is similar to our best-fit log normal distribution, but with a larger low-luminosity tail (as suggested in Fig.~\ref{radio}), it would not impact any of our main conclusions, but would significantly increase the total number of pulsars predicted to reside within the Milky Way's globular cluster system, from $N_{\rm MSP}\sim 50-100$ to roughly $N_{\rm MSP}\sim 100-300$. In reality, this tail might be even larger, depending on the underlying distributions of the parameters $B$ and $\eta_{\gamma}$.

We also note that the luminosity function derived in this study is similar to those found in previous studies for MSPs in the field of the Milky Way. In Fig.~\ref{lzerosigma}, we compare the luminosity function parameters favored by this paper's analysis to that found in Ref.~\cite{Hooper:2015jlu} for those MSPs {\it not} residing in globular clusters. These results overlap at the 1$\sigma$ level, suggesting similar gamma-ray luminosity functions for these two pulsar populations.

Lastly, we point out a potential caveat to the conclusions reached in this section. Thus far, our analysis has assumed that the mean number of MSPs in a given globular cluster is proportional to either the cluster's stellar encounter rate, or to its visible luminosity (or some linear combination of the two). This assumption is both well motivated, and is supported by the observed correlations with gamma-ray luminosity, as shown in Figs.~\ref{GammaE} and~\ref{Vlum}. It is possible, however, that some other variable, perhaps associated with the characteristics of the stellar population, may also be responsible in part for determining the mean number of MSPs within a given cluster. The best-fit model presented here, without any such additional variable(s), provides a fairly good fit to the data shown in Fig.~\ref{GammaE}, at the level of $\chi^2 \simeq 1.08$ per degree-of-freedom. From this, we conclude that the data does not require the inclusion of any additional variable(s). That being said, such a variable could be approximately degenerate with those determined in our fit, and we cannot rule out such a possibility at this time. In the unlikely event that such variables are responsible for a significant fraction of the observed variation in globular cluster luminosities, the parameters favored by our fit would shift to somewhat higher values for $R_{\rm MSP}$ and to lower values for $\langle L_{\gamma} \rangle$.

\section{Disrupted Globular Clusters and the Galactic Center Gamma-Ray Excess}

As discussed in the introduction of this paper, it has been suggested that the gamma-ray excess observed from the region surrounding the Galactic Center might originate from MSPs that were once in globular clusters~\cite{Brandt:2015ula,Bednarek:2013oha}. The results of this study have direct implications for this hypothesis, which we will discuss in this section.

As a consequence of dynamical friction, the orbit of a globular cluster of mass $M_{_{\rm C}}$ will evolve as follows:
\begin{eqnarray}
\frac{dr^2}{dt} = -\frac{r^2}{t_{_{\rm DF}} (r, M_{_{\rm C}})},
\end{eqnarray}
where the timescale for dynamical friction is given by:
\begin{equation}
t_{_{\rm DF}} \approx 4.5 \, {\rm Gyr} \, \times \bigg(\frac{r}{{\rm kpc}}\bigg)^2 \, \bigg(\frac{V(r)}{200 \, {\rm km/s}}\bigg) \, \bigg(\frac{10^6\, M_{\odot}}{M_{_{\rm C}}}\bigg) \, \bigg(\frac{f_{\epsilon}}{0.5}\bigg).
\end{equation}
Here, $V_{\rm C}$ is the circular velocity of the cluster orbit and $f_{\epsilon}$ is an order one factor which accounts for any eccentricity. From this equation, we see that the orbits of massive globular clusters located within the innermost kiloparsecs of the Milky Way will evolve on relevant timescales. As these clusters migrate into the Inner Galaxy, they will lose mass and can be disrupted by tidal forces, causing their pulsars and other stars to be deposited throughout the volume of the Milky Way and especially within the Central Stellar Cluster. In particular, the authors of Ref.~\cite{Gnedin:2013cda} estimate that approximately $1.5 \times 10^8\, M_{\odot}$ in stars has been deposited within the central 1.8 kiloparsecs of the Milky Way from globular clusters, a significant fraction of which is predicted to be concentrated within the innermost several parsecs. This is approximately 3-4 times as much mass as is found within the entirety of the Milky Way's current globular system (see also, Ref~\cite{2016arXiv160605651S}). For a luminosity function with the best-fit parameters found in this study, this would correspond to a total population of MSPs with a gamma-ray luminosity of $L_{\gamma} \simeq (5.2-7.0)\times 10^{36}$ erg/s, corresponding to approximately 27\% to 36\% of the observed gamma-ray excess. As this luminosity function is similar to that reported previously~\cite{Hooper:2015jlu}, this would lead to a similar number of MSPs in the Inner Galaxy ($\sim$~$10-50$) that should have already been detected by Fermi. This estimate, however, neglects the important effects of spin-down evolution, to which we will now turn our attention.

When a given globular cluster is disrupted and its MSP population is deposited into the Inner Galaxy, this population will continue to evolve. More specifically, the pulsars will steadily spin-down, losing their rotational kinetic energy through magnetic dipole breaking and becoming less luminous. This is important for two independent reasons. First, this effect reduces the number of low-mass X-ray binaries (LMXBs) predicted to be present in the Inner Galaxy~\cite{Brandt:2015ula}. And second, this very significantly reduces the average gamma-ray luminosity of MSPs in the Inner Galaxy.

It was first pointed out in Ref.~\cite{Cholis:2014lta} that if the Galactic Center gamma-ray excess is generated by MSPs, then the ratio of MSPs to bright low-mass X-ray binaries (LMXBs) in the Inner Galaxy must be much higher than is observed in globular clusters (by a factor of $\sim$20). As LMXBs are the progenitors of MSPs, this was presented as strong evidence against there being such a large population of MSPs around the Galactic Center. It was later argued, however, that MSPs deposited through the disruption of globular clusters do not need to be accompanied by as large a number of LMXBs~\cite{Brandt:2015ula}. More specifically, when a globular cluster is disrupted, its stellar encounter rate is reduced dramatically, effectively ending the formation of new LMXBs (and ending the formation of new MSPs, other than those that form from preexisting LMXBs). If the timescale for the transition from an LMXB to a MSP is roughly a few hundred million years or less (which is a reasonable estimate~\cite{Humphrey:2006et}), this could explain the lack of bright LMXBs found in the Inner Galaxy~\cite{Revnivtsev:2008fe,Krivonos:2012sd}.  

To estimate the degree to which the gamma-ray luminosity of a given MSP will decrease with time, consider the timescale for spin-down (see Eqs.~\ref{lum} and~\ref{pdot}):   
\begin{eqnarray}
%
\tau \equiv \frac{E}{\dot{E}}= \frac{P}{2\dot{P}} \simeq 0.46 \, {\rm Gyr} \times \bigg(\frac{3\times 10^{34} \, {\rm erg/s}}{L_{\gamma}}\bigg) \, \bigg(\frac{\eta_{\gamma}}{0.2}\bigg) \, \bigg(\frac{3 \, {\rm ms}}{P}\bigg)^2,
\label{pdotrate}
\end{eqnarray}
where $E$ is the rotational kinetic energy of the pulsars. From this, we see that gamma-ray bright MSPs evolve significantly, losing most of their energy over hundreds of millions or billions of years.  As a result, the average gamma-ray luminosity of an MSP from a disrupted globular clusters will be significantly lower than that from an average MSP in the Galactic Disk or within an intact globular cluster (where new pulsars are formed, maintaining something similar to a steady-state luminosity function). In one respect, this might be helpful, as Fermi has failed to detect many bright MSP candidates from the Inner Galaxy (especially in the spherically symmetric configuration required to explain the excess)~\cite{Cholis:2014lta,Lee:2015fea,TheFermi-LAT:2015kwa}, and spin-down evolution might be able to provide an explanation for the lack of high-luminosity sources. After taking into account spin-down evolution, however, much larger numbers of MSPs are required to generate the observed luminosity of the excess.

\begin{figure}
\includegraphics[keepaspectratio,width=0.8\textwidth]{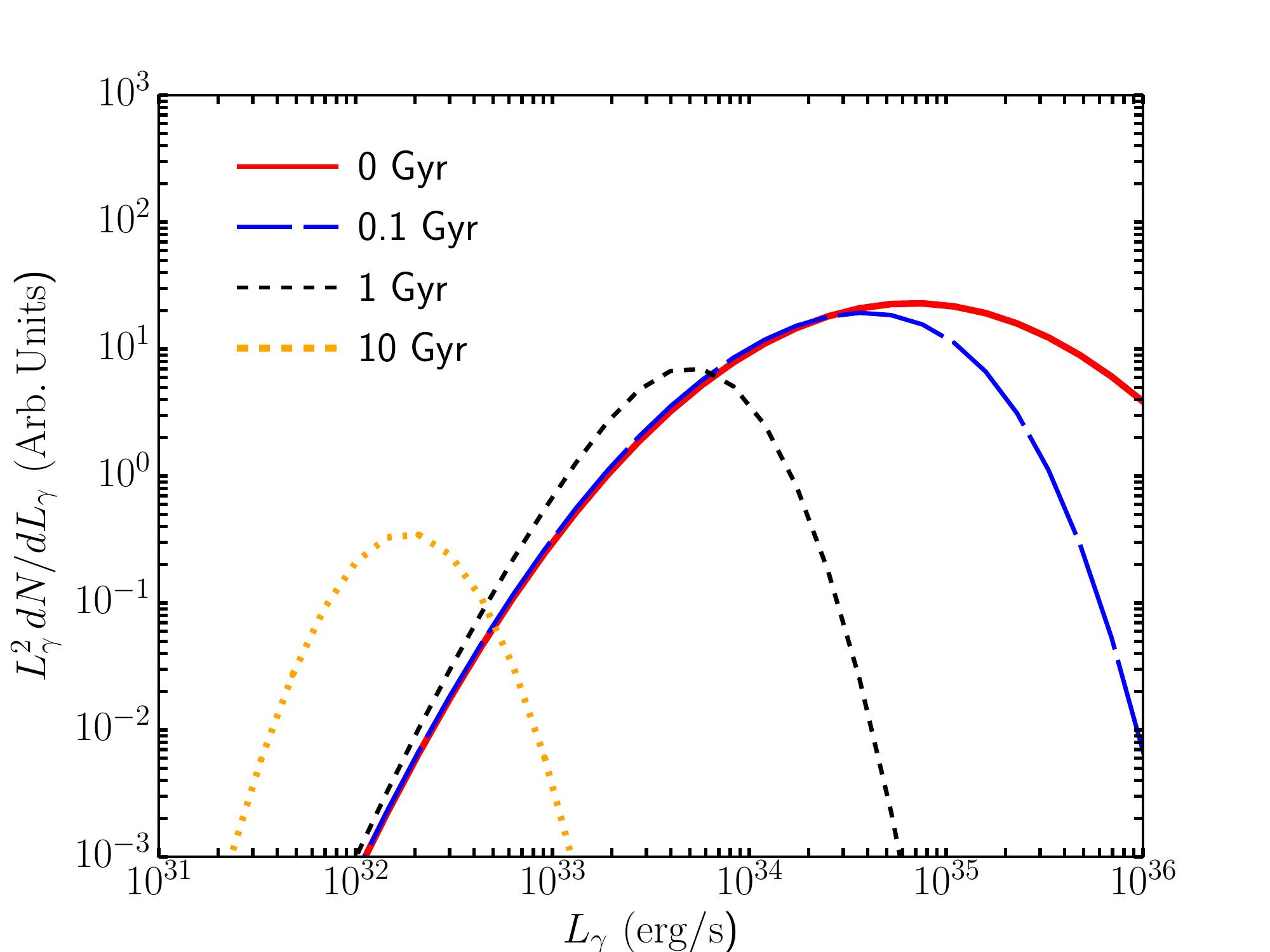}
\caption{The evolution of the millisecond pulsar gamma-ray luminosity function with time. The solid red curve represents the steady-state luminosity function, corresponding to the best-fit parameters found in this study ($L_0=0.88 \times 10^{34}$ erg/s, $\sigma_L=0.62$). The other curves denote the luminosity functions for those same pulsars, after spinning down for a period of 0.1 Gyr (long-dashed blue), 1 Gyr (dashed black) and 10 Gyr (dotted orange). Here, we have assume log-normal distributions for the periods, magnetic field strengths, and gamma-ray efficiencies, with central values of 3 milliseconds, $10^{8.5}$ G, and 0.183, respectively, and each with equal widths (selected such that $\sigma_L=0.62$).}
\label{evolution}
\end{figure}

To estimate the impact of spin-down evolution on a population of MSPs, we have carried out the following calculation. We adopt a distribution of initial periods, magnetic fields and gamma-ray efficiencies that are each described by uncorrelated log-normal distributions, with parameters chosen to recover our best-fit gamma-ray luminosity function ($L_0=0.88 \times 10^{34}$ erg/s, $\sigma_L=0.62$). More specifically, we take $P_0=3$ ms, $B_0=10^{8.5}$ G, and $\eta_{\gamma,0}=0.183$, and $\sigma_P=\sigma_B=\sigma_{\eta_{\gamma}}$ (selected such that $\sigma_L=0.62$). In Fig.~\ref{evolution}, we show the initial luminosity function, and the luminosity function after 0.1, 1.0 and 10.0 Gyr, as calculated using Eq.~\ref{pdotrate}. This calculation reveals that the total gamma-ray luminosity of an MSP population is reduced to 50\% of its initial value after only 0.17 Gyr. After 1.4 Gyr (7.4 Gyr) such a population will emit a total gamma-ray luminosity that is equal to only 10\% (1\%) of its initial value.

\begin{figure}
\includegraphics[keepaspectratio,width=0.8\textwidth]{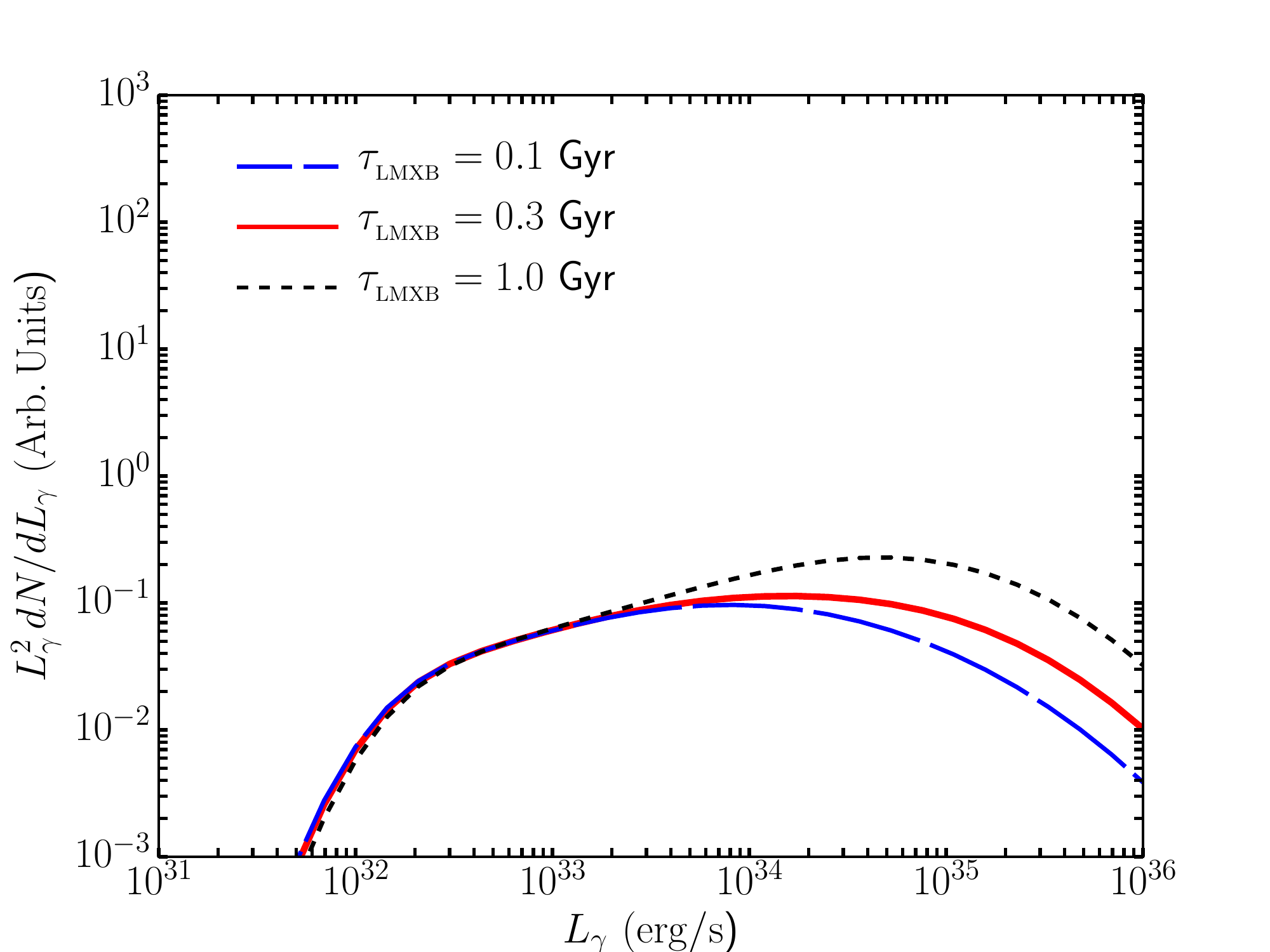}
\caption{The evolution of the gamma-ray luminosity function for the population of millisecond pulsars deposited in the Inner Galaxy from disrupted globular clusters. Results are shown for three different values of the low-mass X-ray binary lifetime, $\tau_{_{\rm LMXB}}$.}
\label{evolution2}
\end{figure}

Taking this a step further, we have assumed that MSPs are deposited from globular clusters into the Inner Galaxy at a constant rate over 10 Gyr (as approximately found to be the case in Ref.~\cite{Gnedin:2013cda}), and calculated the integrated luminosity function of the accumulated MSP population. This result is shown in Fig.~\ref{evolution2}, for three different values of the LMXB lifetime, $\tau_{_{\rm LMXB}}$. For $\tau_{_{\rm LMXB}} \simeq 0.1-0.3$ Gyr~\cite{Humphrey:2006et}, the total
gamma-ray luminosity is a factor of 13 to 17 lower than that found neglecting spin-down evolution. Even for a very large value of  $\tau_{_{\rm LMXB}} \simeq 1.0$ Gyr, the total luminosity is reduced by a factor of 7.4.\footnote{We take the members of the initial LMXB population to have a half-life of $\tau_{_{\rm LMXB}}$. After these sources transition to their MSP phase, their luminosity function undergoes spin-down evolution and departs from the steady-state distribution.}
 
With these considerations in mind, we find that MSPs from disrupted globular clusters are not predicted to generate on the order of 27\% to 36\% of the observed gamma-ray excess, as calculated above, but only 1.6\% to 2.8\% of this signal (for $\tau_{_{\rm LMXB}} \simeq 0.1-0.3$ Gyr). Furthermore, even if we were to radically change the assumed initial conditions of the globular cluster population such that 30 to 70 times more MSPs from globular clusters were deposited into the Inner Galaxy, this would lead to the formation of a central stellar population that is much larger than observed ($\sim$~$10^9 \,  M_{\odot}$ within the innermost 10 pc, compared to the observed value of $3\times 10^{7} \, M_{\odot}$)~\cite{Gnedin:2013cda}. In such a scenario, globular clusters would also contribute a total stellar mass of $\sim10^{10} \, M_{\odot}$ within the innermost 1.8 kpc of the Galaxy, constituting an order one fraction of total mass of the bulge.

One caveat to the conclusions described above concerns the population
of MSPs that is ejected from the globular cluster by the natal kick
imparted during the supernova of the primary star. These systems may
still become MSPs, but the emission will not be localized in the
globular cluster. Within the context of our analysis, this will lead to an observed gamma-ray signal from the globular cluster population that underestimates the total gamma-ray luminosity from pulsars born
within that cluster. There are several reasons, however, that we expect this correction to be small. First of all, dynamical simulations indicate that only $\sim$10\% of neutron stars that form and retain their binary companion in a globular cluster are expelled from the cluster~\cite{1998MNRAS.301...15D}. Furthermore, a comparison of the $\sim$Gyr timescale for dynamical interactions in globular clusters~\cite{Fregeau:2009mz} to the $\sim$20~Myr timescale for the formation of a neutron star indicates that the overwhelming majority of MSPs in globular clusters originate from primordial binaries, rather than from systems with a captured companion. This class of systems is not expected to have an MSP formation efficiency that is enhanced with respect to the field ({\ie}the thick disk).\footnote{While Globular
Clusters are likely to support a larger initial binary fraction than
stars in the field, the binary fraction of stars in the field is
likely to be $\sim$0.5, limiting any enhancement to be
$\lsim$2~\cite{Goodwin:2009vm}.} And lastly, we note that globular clusters spend much of their pre-disruption lifetime in orbits well outside of region relevant for studies of the Galactic Center gamma-ray excess. For example, following Ref.~\cite{Gnedin:2013cda}, we find that a cluster with an initial mass of $10^7 \, M_{\odot}$ and in an orbit with a semi-major axis of 4 kpc will migrate into the Inner Galaxy over a period of approximately 8.6 Gyr. Only about 28\% of this time is spent within the innermost 2 kpc of the Galaxy, however. Less massive clusters migrate even more slowly. As a consequence, a large fraction of MSPs expelled from globular clusters through natal kicks will be found well beyond the confines of the Inner Galaxy.

\section{Summary and Conclusions}

Millisecond pulsars (MSPs) are widely considered to be the leading astrophysical explanation for the gamma-ray excess observed from the region surrounding the Galactic Center. The morphology of this gamma-ray signal is approximately spherically symmetric, however, and does not trace any known stellar populations, making it difficult to explain with any pulsar population that might be generated within the Inner Galaxy. With this in mind, it has been proposed that a large and more spherically distributed population of MSPs might have originated from the tidal disruption of globular clusters.

In this paper, we have studied the gamma-ray emission from 157 globular clusters in the Milky Way. Within this population, we have identified statistically significant emission from 25 globular clusters, 10 of which are not found in existing gamma-ray source catalogs. We then made use of this information, in conjunction with previously calculated stellar encounter rates, to constrain the gamma-ray luminosity function of MSPs in globular clusters. Our results suggest that this population of MSPs exhibit a luminosity function that is quite similar to that of MSPs in the field of the Milky Way ({\ie}in the thick disk).

After taking into account the spin-down evolution of MSPs deposited into the Inner Galaxy, we find that it would require between 3300 and 4200 MSPs (with gamma-ray luminosities $L_{\gamma} > 10^{33}$ erg/s) to generate the observed intensity of the gamma-ray excess (for our best-fit luminosity function parameters). If we adopt the luminosity function derived in this study, the Milky Way globular cluster model described in Refs.~\cite{Gnedin:2013cda,Brandt:2015ula} predicts a total gamma-ray flux that is only a 1.6\% to 2.8\% as large as the observed excess. A scenario that deposited enough MSPs into the Inner Galaxy to generate the entire gamma-ray excess would also lead to the formation of a $\sim$$10^{9}\, M_{\odot}$ central stellar cluster within the innermost 10 pc, a factor of $\sim$30 more mass than is observed to be present. If we change the parameters of the spin-down evolution model in order to generate a larger total gamma-ray flux, this leads to the prediction that Fermi should have already detected many MSPs in the Inner Galaxy, as well as many more bright low-mass X-ray binaries than are observed.

\bigskip

\textbf{Acknowledgments.} We would like to thank Roland Crocker, Alex Drlica-Wagner, and David Nataf for helpful discussions. DH is supported by the US Department of Energy under contract DE-FG02-13ER41958. Fermilab is operated by Fermi Research Alliance, LLC, under Contract No. DE-AC02-07CH11359 with the US Department of Energy. TL is supported by the National Aeronautics and Space Administration through Einstein Postdoctoral Fellowship Award No. PF3-140110. We acknowledge the Ohio Supercomputer Center for providing support for this work.

\bibliography{globularclusters.bib}
\bibliographystyle{JHEP}

\end{document}